\documentclass[aps,prc,superscriptaddress,twocolumn]{revtex4}
\usepackage{latexsym}
\usepackage{amssymb}
\usepackage{amsmath}
\usepackage{graphicx}
\usepackage{bm}
\usepackage{epsf}
\usepackage{epstopdf}
\usepackage{color,xcolor}
\usepackage{ulem}

\begin{document}

\title{Hadron-quark mixed phase in the quark-meson coupling model}

\author{Min Ju}
\affiliation{School of Physics, Nankai University, Tianjin 300071, China}
\author{Xuhao Wu}
\affiliation{School of Physics, Peking University, Beijing 100871, China}
\author{Fan Ji}
\affiliation{School of Physics, Nankai University, Tianjin 300071, China}
\author{Jinniu Hu}~\email{hujinniu@nankai.edu.cn}
\affiliation{School of Physics, Nankai University, Tianjin 300071, China}
\author{Hong Shen}~\email{shennankai@gmail.com}
\affiliation{School of Physics, Nankai University, Tianjin 300071, China}

\begin{abstract}
We explore the possibility of a structured hadron-quark mixed phase forming in the
interior of neutron stars. The quark-meson coupling (QMC) model, which explicitly
incorporates the internal quark structure of the nucleon, is employed to describe
the hadronic phase, while the quark phase is described by the same bag model
as the one used in the QMC framework, so as to keep consistency between the
two coexisting phases. We analyze the effect of the appearance of hadron-quark
pasta phases on the neutron-star properties.
We also discuss the influence of nuclear symmetry energy and the bag
constant $B$ in quark matter on the deconfinement phase transition.
For the treatment of the hadron-quark mixed phase, we use the energy minimization
method and compare it with the Gibbs construction.
The finite-size effects like surface and Coulomb energies are taken into
account in the energy minimization method;
they play crucial roles in determining the pasta configuration during the
hadron-quark phase transition. It is found that the finite-size effects can
significantly reduce the region of the mixed phase relative to that of the
Gibbs construction. Using a consistent value of $B$ in the QMC model
and quark matter, we find that hadron-quark pasta phases are formed
in the interior of massive stars, but no pure quark matter can exist.
\end{abstract}

\maketitle

\section{Introduction}
\label{sec:1}

It is generally believed that hadronic matter undergoes a phase transition
to deconfined quark matter at high baryon densities, which may occur in
the interior of massive neutron stars~\cite{Baym18,Glen01,Webe05}.
Recent advances in astrophysical observations have led to increasing interest
in exploring various properties of neutron stars, such as their internal
structure, mass-radius relation, and tidal deformability.
Several precise mass measurements of massive pulsars,
PSR J1614-2230~\citep{Demo10,Fons16,Arzo18}, PSR J0348+0432~\citep{Anto13},
and PSR J0740+6620~\citep{Crom19}, constrain the maximum neutron-star mass
to be larger than about $2 M_\odot$, which challenges our understanding of the
equation of state (EOS) of superdense matter.
The first detection of gravitational waves from a binary neutron-star merger,
known as GW170817, provides valuable constraints on the tidal
deformability~\citep{Abbo17,Abbo18,Abbo19}, which also restricts the radii of
neutron stars~\cite{Tews18,De18,Fatt18,Mali18,Zhu18}.
More recently, the gravitational-wave event GW190425 was
reported by LIGO and Virgo Collaborations~\citep{Abbo190425} with the total
mass of the binary system as large as $3.4^{+0.3}_{-0.1} M_\odot$,
which may offer important information for the EOS at high densities.
The latest gravitational-wave event GW190814 was detected from the merger of
binary coalescence involving a $23.2^{+1.1}_{-1.0} M_\odot$ black hole
with a $2.59^{+0.08}_{-0.09} M_\odot$ compact object~\citep{Abbo190814},
where the secondary component could be interpreted as either
the lightest black hole or the heaviest neutron star ever observed.
Furthermore, the new observations
by the {\it Neutron Star Interior Composition Explorer} ({\it NICER})
provided a simultaneous measurement of the mass and radius for
PSR J0030+0451~\citep{Rile19,Mill19}.
It is interesting to notice that the constraints on the neutron-star radius
from various observations are consistent with each other, and suggest
relatively small radii of neutron stars.
All of these exciting developments in astrophysical observations
provide a wealth of information about neutron star interiors,
where the deconfinement phase transition is expected to take place.

Theoretically, a hadron-quark mixed phase is predicted to exist in the region
between hadronic matter and quark matter based on various approaches~\cite{Glen92,
Sche99,Sche00,Latt00,Burg02,Mene03,Yang08,Xu10,Chen13,Orsa14,Wu17,Baym18}.
However, large uncertainties in the structure and density range of the mixed phase
are present due to the different models and methods used.
The Gibbs construction~\cite{Glen92} is generally adopted for the description
of hadron-quark mixed phase, where the coexisting hadronic and quark phases
are allowed to be charged separately, but the finite-size effects like surface
and Coulomb contributions are neglected.
When the surface tension of the hadron-quark interface is sufficiently large,
the mixed phase should be described by the Maxwell construction,
where local charge neutrality is imposed and coexisting hadronic and quark
phases have equal pressures and baryon chemical potentials but different
electron chemical potentials.
It is evident that Gibbs and Maxwell constructions correspond to two limits of
vanishing and large values of the surface tension, respectively, while
both of them involve only bulk contributions without finite-size effects~\cite{Bhat10,Wu19}.
For a moderate surface tension, the hadron-quark mixed phase with some geometric
structures, known as pasta phases, is expected to appear as a consequence of
the competition between the surface and Coulomb
energies~\cite{Wu19,Heis93,Endo06,Maru07,Yasu14,Spin16}.
A realistic description of the structured mixed phase was proposed in
Refs.~\cite{Tats03,Endo06}, where a consistent treatment of the electric field in
the Wigner--Seitz cell would lead to inhomogeneous distributions of charged
particles in both hadronic and quark phases. In general, for simplicity,
the particle densities in the two coexisting phases are assumed to be
spatially constant, and the charge screening effect is neglected.
A simple coexisting phases (CP) method~\cite{Wu19} is used for the
description of the hadron-quark pasta phases, where the coexisting phases
satisfy the Gibbs conditions for phase equilibrium, and the finite-size effects
like the surface and Coulomb energies are perturbatively taken into account.
An improved energy minimization (EM) method incorporates the finite-size effects
in a more consistent manner~\cite{Wu19}, where the equilibrium conditions for coexisting
phases are derived by minimization of the total energy including surface and
Coulomb contributions. As a result, the equilibrium conditions obtained in
the EM method are significantly different from the Gibbs conditions.
The EM method, known as the compressible liquid-drop (CLD) model,
has been widely used in the study of nuclear liquid-gas phase transition at
subnuclear densities~\cite{Bao14b,Baym71,Latt85,Latt91}.
In the present work, we intend to use the EM method for studying the hadron-quark
pasta phases in the interior of massive neutron stars.

To describe the hadron-quark pasta phases, we employ the Wigner--Seitz approximation,
in which the system is divided into equivalent and charge-neutral cells
with a given geometric symmetry.
The hadronic and quark phases inside the cell are assumed to have constant densities
and are separated by a sharp interface.
For the description of hadronic matter, we employ the quark-meson coupling (QMC)
model, which explicitly incorporates the quark degrees of freedom.
The QMC model, initially proposed by Guichon~\cite{Guic88},
has been extensively developed and applied to various nuclear and
hadronic phenomena in the past decades~\cite{Sait96,Mull98,Yue06,Sait07,Guic18,Gram17,Mott19}.
In the QMC model, the nucleons in the nuclear medium are described by static MIT bags
that interact through the self-consistent exchange of scalar and vector mesons in the
mean-field approximation. The mesons couple directly to the confined quarks
inside the bags, not to pointlike nucleons as in the relativistic mean-field (RMF) model.
In contrast to the RMF approach, the internal structure of the nucleon is explicitly
included in the QMC model, so it could be used for investigating the medium
modification of nucleon structure~\cite{Guic18}.
At sufficiently high densities, the nucleons and other hadrons are expected to dissolve
into quarks. The quark matter is then treated as a
degenerate Fermi gas of $u$, $d$, and $s$ quarks within the MIT bag model.
In this sense, both hadronic and quark phases are consistently described
within the same bag model. Therefore, the QMC model that incorporates
the internal quark structure of the nucleon is a promising approach for
exploring the hadron-quark phase transition in neutron stars.
In Refs.~\cite{Panda04a,Panda04b,Panda10,Panda12}, Panda \textit{et al.}
extended the QMC model to study the properties of compact stars,
including the possibility of hyperon formation and quark deconfinement.
They employed the Gibbs construction to describe the hadron-quark mixed phase
without considering finite-size effects and possible geometric structures.
In the present work, we aim to study the hadron-quark pasta phases using
the QMC model, where the finite-size effects are considered within the EM method,
and the properties of a structured mixed phase are investigated.

This article is organized as follows.
In Sec.~\ref{sec:2}, we describe the QMC model for the hadronic matter.
In Sec.~\ref{sec:3}, the MIT bag model for the quark matter is briefly reviewed.
The EM method for the description of hadron-quark pasta phases
is presented in Sec.~\ref{sec:4}, where the surface tension at
the hadron-quark interface is discussed.
Section~\ref{sec:5} contains numerical results of the hadron-quark pasta phases
and corresponding neutron-star properties.
Section~\ref{sec:6} is devoted to the conclusions.

\section{Hadronic phase}
\label{sec:2}
The hadronic matter is described by the QMC model, where the nucleon is
treated as a static MIT bag containing three confined quarks.
The nucleon-nucleon interaction is realized by the exchange of isoscalar-scalar
meson $\sigma$, isoscalar-vector meson $\omega$, and isovector-vector meson $\rho$
in the mean-field approximation. In the QMC model, the effective meson fields
couple directly to the confined quarks inside the nucleon, not to the pointlike
nucleon as in the RMF model. Therefore, the internal structure of the nucleon
can be self-consistently determined, and is influenced by the meson fields
in nuclear matter. The quark field inside the bag satisfies the Dirac equation
\begin{equation}
\label{2.1}
\left[  i\gamma^{\mu}\partial_{\mu}-\left(m_{q}+g_{\sigma}^{q}\sigma\right)
-\gamma^{0}\left(g_{\omega}^{q}\omega+g_{\rho}^{q}\tau_{3}\rho\right)\right]\psi_{q}=0,
\end{equation}
where $m_{q}$ is the current quark mass and $g_{\sigma}^{q}$, $g_{\omega}^{q}$,
and $g_{\rho}^{q}$ denote the quark-meson coupling constants.
The normalized ground state for a quark in the bag is given by
\begin{equation}
\label{2.2}
\psi_{q}({\bf r},t)={\cal N}_{q}e^{-i\epsilon_{q}t/R}\left( \begin{array}{c}
j_{0}(x_{q}r/R)  \\
i\beta_{q} {\bf \sigma}\cdot\hat{\bf r}j_{1}(x_{q}r/R)
\end{array}\right) \frac{\chi_{q}}{\sqrt{4\pi}},
\end{equation}
where
\begin{eqnarray}
\label{2.3}
\epsilon_{q}&=&\Omega_{q}+R\left(g_{\omega}^{q}\omega+g_{\rho}^{q}\tau_{3}\rho\right),\\
\beta_{q}&=&\sqrt{\frac{\Omega_{q}-Rm_{q}^{*}}{\Omega_{q}+Rm_{q}^{*}}},   \\
{\cal N}_{q}^{-2}&=& 2R^{3}j_{0}^{2}(x_{q})[\Omega_{q}(\Omega_{q}-1)+Rm_{q}^{*}/2]/x_{q}^{2},
\end{eqnarray}
with $\Omega_{q}=\sqrt{x_{q}^{2}+(Rm_{q}^{*})^{2}}$ and $m_{q}^{*}=m_{q}+g_{\sigma}^{q}\sigma$.
Here $R$ is the bag radius and $\chi_{q}$ is the quark spinor. The value of $x_{q}$ is
determined by the boundary condition at the bag surface
\begin{equation}
j_{0}(x_{q})=\beta_{q}j_{1}(x_{q}).
\end{equation}
The energy of a nucleon bag consisting of three ground state quarks is then given by
\begin{equation}\label{2.5}
E_{\rm{bag}}=3\frac{\Omega_{q}}{R}-\frac{Z}{R}+\frac{4}{3}\pi R^{3}B,
\end{equation}
where the model parameter $Z$ accounts for the zero-point motion and center-of-mass corrections,
and $B$ denotes the bag constant. The effective nucleon mass in the medium is taken to be
$M_{N}^{*}=E_{\rm{bag}}$. The bag radius $R$ is determined by the equilibrium condition
$\partial M_{N}^{*}/\partial R = 0$.
In the present work, we use the current quark mass $m_{q}=5.5$~MeV for the $u$ and $d$ quarks.
The model parameters, $B^{1/4}=210.854$~MeV and $Z=4.00506$, are determined by reproducing
the nucleon mass $M_{N}=939$~MeV and the bag radius $R=0.6$ fm in free space~\cite{Panda04a, Yue06}.

To describe the hadronic matter consisting of nucleons ($p$ and $n$) and
leptons ($e$ and $\mu$), we start with the effective Lagrangian density involving
the internal structure of the nucleon together with the meson fields
in the mean-field approximation,
\begin{eqnarray}
\label{2.6}
\mathcal{L_{\mathrm{QMC}}} &=& \sum_{b=n,p}\bar{\psi}_{b}
\left[ i\gamma _{\mu}\partial^{\mu }-M_{N}^{\ast }
-\gamma^{0}\left(g_{\omega}\omega+g_{\rho}\tau_{3}\rho\right)\right]\psi_{b}
\nonumber\\
&&-\frac{1}{2}m_{\sigma }^{2}\sigma ^{2}+\frac{1}{2}m_{\omega }^{2}\omega ^{2}
+\frac{1}{2}m_{\rho}^{2}\rho^{2}
+\Lambda_{\rm{v}}g_{\omega}^{2}g_{\rho}^{2}\omega^{2}\rho^{2}  \nonumber\\
&&+\sum_{l=e,\mu}\bar{\psi}_{l}[i\gamma_{\mu}\partial ^{\mu}-m_{l}]\psi_{l},
\end{eqnarray}
where $\sigma =\left\langle \sigma \right\rangle$,
$\omega =\left\langle \omega^{0}\right\rangle$,
and $\rho =\left\langle \rho^{30} \right\rangle$ are the nonvanishing expectation values
of meson fields in homogeneous nuclear matter.
The effective nucleon mass $M_{N}^{\ast}$ is calculated from
Eq.~(\ref{2.5}) at the quark level, and it depends on the $\sigma$ field.
Generally, the effective nucleon mass can be expanded in terms of $\sigma$,
which is written in the practical form:
\begin{equation}
\label{2.7}
M_{N}^{*}=M_{N}+a\left(g_{\sigma}^{q}\sigma\right)
          +b \left(g_{\sigma}^{q}\sigma\right)^{2}
          +c \left(g_{\sigma}^{q}\sigma\right)^{3},
\end{equation}
where the parameters $a=1.45162$, $b=7.75404\times10^{-4}$~MeV$^{-1}$, and
$c=1.38043\times10^{-7}$~MeV$^{-2}$ are determined by fitting to the results
of $M_{N}^{\ast}$ from Eq.~(\ref{2.5}).
In the QMC model, the fundamental couplings are those to the quarks,
which are related to the corresponding nucleon-meson couplings
as $g_{\omega}=3g_{\omega}^{q}$ and $g_{\rho}=g_{\rho}^{q}$.
In practical calculations, we use the quark-meson coupling
constants $g_{\sigma}^{q}=5.9895$, $g_{\omega}^{q}=3.0018$,
and $g_{\rho}^{q}=5.6013$, together with the $\omega$-$\rho$
coupling $\Lambda_{\rm{v}}=0.0844$. These parameters are obtained
by reproducing the nuclear matter properties of binding energy $E/A=-16.3$~MeV
at the saturation density $n_{0}=0.15$ fm$^{-3}$ with the symmetry energy
$E_{\rm{sym}}=31$~MeV and its slope $L=40$~MeV.
It is well known that the $\omega$-$\rho$ coupling term plays a crucial role
in determining the density dependence of the symmetry energy~\cite{Bao14b}.
We include this coupling term in Eq.~(\ref{2.6}), so that a small value of the
slope $L$ could be obtained according to current constraints.
The meson masses are taken to be $m_{\sigma}=550$~MeV, $m_{\omega}=783$~MeV,
and $m_{\rho}=770$~MeV.

In uniform hadronic matter, the coupled equations of motion for meson fields can be
easily solved. The total energy density of hadron matter is calculated by
\begin{eqnarray}
\varepsilon_{\rm{HP}} &=&\sum_{b=p,n}\frac{1}{\pi^2}
     \int_{0}^{k^{b}_{F}}{\sqrt{k^2+{M_N^{\ast}}^2}}\ k^2dk   \nonumber \\
&& + \frac{1}{2}m^2_{\sigma}{\sigma}^2 + \frac{1}{2}m^2_{\omega}{\omega}^2
   + \frac{1}{2}m^2_{\rho}{\rho}^2
   +3\Lambda_{\rm{v}}g_{\omega}^{2}g_{\rho}^{2}\omega^{2}\rho^{2}
     \nonumber  \\
&& + \sum_{l=e,\mu}\frac{1}{\pi^{2}}\int_{0}^{k_{F}^{l}}
     \sqrt{k^{2}+m_{l}^{2}}\ k^{2}dk,
\label{eq:ehp}
\end{eqnarray}
and the pressure is given by
\begin{eqnarray}
P_{\rm{HP}} &=& \sum_{b=p,n}\frac{1}{3\pi^2}\int_{0}^{k^{b}_{F}}
      \frac{k^4dk}{\sqrt{k^2+{M_N^{\ast}}^2}}   \nonumber  \\
&& - \frac{1}{2}m^2_{\sigma}{\sigma}^2 + \frac{1}{2}m^2_{\omega}{\omega}^2
   + \frac{1}{2}m^2_{\rho}{\rho}^2
   + \Lambda_{\rm{v}}g_{\omega}^{2}g_{\rho}^{2}\omega^{2}\rho^{2} \nonumber \\
&& + \sum_{l=e,\mu}\frac{1}{3\pi^{2}}\int_{0}^{k_{F}^{l}}
     \frac{k^{4} dk}{\sqrt{k^{2}+m_{l}^{2}}}.
\label{eq:php}
\end{eqnarray}
For hadronic matter consisting of nucleons and leptons in $\beta$ equilibrium,
the chemical potentials satisfy the relations
\begin{eqnarray}
\mu_{p} &=& \mu_{n}-\mu_{e}, \\
\mu_{\mu}&=& \mu_{e}.
\end{eqnarray}
The charge neutrality condition is expressed as
\begin{eqnarray}
n_{c}^{\mathrm{HP}}=n_{p}-n_{e}-n_{\mu}=0.
\end{eqnarray}
It is known that muons may appear when the electron chemical potential
exceeds the muon mass, which occurs at a density of $0.11$ fm$^{-3}$
in the QMC model.

\section{Quark phase}
\label{sec:3}

To describe the quark matter at high densities, we employ the same bag model
as the one used to describe the nucleon in the QMC model.
In its simplest form, the quarks are treated as a noninteracting Fermi gas confined
in a large bag, where the zero-point energy characterized by the parameter $Z$ is neglected.
We consider the quark matter consisting of three flavor quarks ($u$, $d$, and $s$) and
leptons ($e$ and $\mu$) in $\beta$ equilibrium, where the relations between chemical
potentials are expressed as
\begin{eqnarray}
\mu_{s} &=& \mu_{d}=\mu_{u}+\mu_{e}, \label{eq:qchq}\\
\mu_{\mu}&=& \mu_{e}.   \label{eq:qche}
\end{eqnarray}
The charge neutrality condition is written as
\begin{eqnarray}
n_{c}^{\mathrm{QP}}=\frac{2}{3}n_{u}-\frac{1}{3}n_{d}-\frac{1}{3}n_{s}-n_{e}-n_{\mu}=0.
\label{eq:qnc}
\end{eqnarray}
We use the current quark masses $m_u=m_d=5.5$~MeV and $m_s=150$~MeV in the calculation
of quark matter, which is consistent with that used in the QMC model.
As for the bag constant, we mainly take the value $B^{1/4}=210.854$~MeV determined in
the QMC model. In fact, the value of $B$ in quark matter may be different from
the one used to describe the nucleon in the QMC model due to the large density
difference between the two cases.
Therefore, we will compare results with different choices of $B$ for quark matter,
so as to examine the influence of the bag constant.

For quark matter described in the MIT bag model, the energy density and pressure
are given by
\begin{eqnarray}
\varepsilon_{\rm{QP}} &=&\sum_{i=u,d,s}\frac{3}{\pi^2}
     \int_{0}^{k^{i}_{F}}{\sqrt{k^2+m_i^2}}\ k^2dk +B  \nonumber  \\
 &&  + \sum_{l=e,\mu}\frac{1}{\pi^{2}}\int_{0}^{k_{F}^{l}}
     \sqrt{k^{2}+m_{l}^{2}}\ k^{2}dk,
\label{eq:eqp}
\end{eqnarray}
\begin{eqnarray}
P_{\rm{QP}} &=& \sum_{i=u,d,s}\frac{1}{\pi^2}\int_{0}^{k^{i}_{F}}
     \frac{k^4dk}{\sqrt{k^2+m_i^2}} - B  \nonumber  \\
 &&  + \sum_{l=e,\mu}\frac{1}{3\pi^{2}}\int_{0}^{k_{F}^{l}}
     \frac{k^{4} dk}{\sqrt{k^{2}+m_{l}^{2}}},
\label{eq:pqp}
\end{eqnarray}
where the bag constant $B$ is added to the energy density and
subtracted from the pressure. It is well known that the bag constant
could significantly affect the EOS of quark matter and consequently
influence the hadron-quark phase transition in neutron stars~\cite{Sche00}.

\section{Mixed phase within the energy minimization method}
\label{sec:4}
For the description of a hadron-quark mixed phase, we take into account the
finite-size effects by using the Wigner--Seitz approximation in our calculations.
In the Wigner--Seitz approximation, the system is divided into equivalent and
charge-neutral cells, where the coexisting hadronic and quark phases are separated
by a sharp interface with finite surface tension.
The leptons are assumed to be uniformly distributed throughout the cell.
In the EM method, the equilibrium conditions between coexisting
phases are determined by minimization of the total energy including surface and
Coulomb contributions. Due to the competition between surface and Coulomb energies,
the geometric structure of the mixed phase may change from droplet to rod,
slab, tube, and bubble with increasing baryon density; these are known as pasta phases.
The total energy density of the mixed phase is given by
\begin{eqnarray}
\label{eq:fws}
\varepsilon_{\rm{MP}} &=& \chi \varepsilon_{\rm{QP}}
  + \left( 1 - \chi \right)\varepsilon_{\rm{HP}}
  + \varepsilon_{\rm{surf}} + \varepsilon_{\rm{Coul}} ,
\end{eqnarray}%
where $\chi=V_{\rm{QP}}/(V_{\rm{QP}}+V_{\rm{HP}})$ denotes the volume fraction
of the quark phase. The energy densities, $\varepsilon_{\rm{HP}}$ and $\varepsilon_{\rm{QP}}$,
are calculated from Eqs.~(\ref{eq:ehp}) and (\ref{eq:eqp}), respectively.
The first two terms of Eq.~(\ref{eq:fws}) represent the bulk contributions,
while the last two terms come from the finite-size effects.
The surface and Coulomb energy densities are given by
\begin{eqnarray}
{\varepsilon}_{\rm{surf}}
&=& \frac{D \sigma \chi_{\rm{in}}}{r_D},
\label{eq:esurf} \\
{\varepsilon}_{\textrm{Coul}}
&=& \frac{e^2}{2}\left(\delta n_c\right)^{2}r_D^{2} \chi_{\rm{in}}\Phi\left(\chi_{\rm{in}}\right),
\label{eq:ecoul}
\end{eqnarray}%
with%
\begin{eqnarray}
\label{eq:Du}
\Phi\left(\chi_{\rm{in}}\right)=\left\{
\begin{array}{ll}
\frac{1}{D+2}\left(\frac{2-D\chi_{\rm{in}}^{1-2/D}}{D-2}+\chi_{\rm{in}}\right),  & D=1,3, \\
\frac{\chi_{\rm{in}}-1-\ln{\chi_{\rm{in}}}}{D+2},  & D=2. \\
\end{array} \right.
\end{eqnarray}%
Here $D=1,2,3$ is the geometric dimension of the cell with $r_D$ being the size
of the inner phase. $\chi_{\rm{in}}$ represents the volume fraction of the inner phase,
i.e., $\chi_{\rm{in}}=\chi$ for droplet, rod, and slab configurations,
and $\chi_{\rm{in}}=1-\chi$ for tube and bubble configurations.
$e=\sqrt{4\pi/137}$ is the electromagnetic coupling constant.
$\delta n_c=n_c^{\rm{HP}}-n_c^{\rm{QP}}$ is the charge-density
difference between hadronic and quark phases.
$\sigma$ denotes the surface tension at the hadron-quark interface,
which plays a key role in determining the structure of the mixed phase~\cite{Wu19,Heis93,Endo06,Maru07,Yasu14,Spin16}.
However, considerable uncertainties exist regarding the value of $\sigma$,
so it is often treated as a free parameter in the literature.
Estimates based on different models suggest relatively small values
$\sigma < 30$~MeV/fm$^2$~\cite{Berg87,Palh10,Pint12,Lugo17,Lugo19},
but much larger values $\sigma > 100$~MeV/fm$^2$ have also been
obtained in Refs.~\cite{Alfo01,Lugo13}.

In the present work, we prefer to calculate the surface tension $\sigma$
in the MIT bag model by using the multiple reflection
expansion (MRE) method~\cite{Berg87}.
The derivation of $\sigma$ within the MRE framework is based on the
modified density of states for the quark species $i$ in an MIT bag,
which is approximately given by
\begin{eqnarray}
\frac{d N_i}{d k_i}=6\left[ \frac{k_i^{2} V}{2\pi^{2}}
  -\frac{k_i S}{8\pi}\left(1-\frac{2}{\pi} \arctan \frac{k_i}{m_{i}}\right)\right],
\end{eqnarray}
where $V$ and $S$ are the volume and surface area of the bag, respectively.
According to the relation between the surface tension and the thermodynamic
potential at zero temperature, one can calculate the surface tension contributed
from the quark species $i$ by
{\small \begin{eqnarray}
\sigma_i &=& \int_{0}^{k^{b}_{F}}
\frac{3k_i}{4\pi}\left(1-\frac{2}{\pi} \arctan \frac{k_i}{m_{i}}\right)
\left[\mu_{i}-\sqrt{k_i^2+m_i^2} \right]dk_i \nonumber \\
&=& \frac{3}{4\pi}\left\{ \frac{ {k_{F}^{i}}^{2}\mu_{i}}{6}
   -\frac{m_{i}^{2}\left( {\mu _{i}-m_{i}}\right)}{3}
   -\frac{1}{3\pi} \left[ \mu_{i}^{3}\mathrm{arctan}\frac{k_{F}^{i}}{m_{i}}
   \right. \right. \nonumber \\
& & \left. \left. -2k_{F}^{i}m_{i}\mu_{i}
    +m_{i}^{3} \ln \left(\frac{k_{F}^{i}+\mu_{i}}{m_{i}}\right)  \right] \right\} ,
\label{eq:sigma}
\end{eqnarray} }
where $\mu_{i}=\sqrt{{k_{F}^{i}}^{2}+m_i^2}$ is the chemical potential of quarks.
In the MRE method, the surface tension of the bag is given by a sum over
all flavors, $\sigma=\sum_{i=u,d,s}\sigma_i$. Note that the dominant contribution
to the surface tension comes from the $s$ quark, since its mass is much larger
than those of $u$ and $d$ quarks. On the other hand, $\sigma$ is also dependent
on the density of quark matter.

At a given baryon density, the thermodynamically stable state is the one with
the lowest energy among all configurations considered, which could be determined
in the EM method by minimizing the total energy with respect to all variables.
The energy density of the mixed phase given in Eq.~(\ref{eq:fws})
is considered as a function of the following variables:
$n_{p}$, $n_{n}$, $n_{u}$, $n_{d}$, $n_{s}$, $n_{e}$, $n_{\mu}$, $\chi$, and $r_D$.
The minimization should be performed under the constraints of global charge neutrality
and baryon number conservation, which are expressed as
\begin{eqnarray}
\label{eq:nc}
n_e + n_\mu -\frac{\chi}{3}\left( 2n_u - n_d - n_s \right)
  - \left(1 - \chi\right)  n_p = 0  , \\
\frac{\chi}{3}\left( n_u + n_d + n_s \right)
    + \left(1 - \chi\right) \left( n_p + n_n \right) = n_b . \,
\label{eq:nb}
\end{eqnarray}
We introduce the Lagrange multipliers $\mu_e$ and $\mu_n$
for the constraints, and then perform the minimization for
the function
{\small \begin{eqnarray}
w &=&\varepsilon_{\rm{MP}}
  -\mu_e \left[ n_e + n_\mu - \frac{\chi}{3}(2n_u - n_d - n_s)
    - \left(1 - \chi\right)  n_p  \right] \notag \\
  & & -\mu_n \left[ \frac{\chi}{3}\left( n_u + n_d + n_s \right)
  + \left(1 - \chi\right) \left( n_p + n_n \right)\right].
\end{eqnarray}}
By minimizing $w$ with respect to the particle densities, we obtain
the following equilibrium conditions for chemical potentials:
\begin{eqnarray}
\mu_u - \frac{ 4\varepsilon_{\rm{Coul}} }{3\chi \, \delta n_c}
  &=& \frac{1}{3}\mu_n - \frac{2}{3}\mu_e, \label{eq:CU1}\\
\mu_d + \frac{ 2\varepsilon_{\rm{Coul}} }{3\chi \, \delta n_c}
  &=& \frac{1}{3}\mu_n + \frac{1}{3}\mu_e, \label{eq:CD1}\\
\mu_s + \frac{ 2\varepsilon_{\rm{Coul}} }{3\chi \, \delta n_c}
  &=& \frac{1}{3}\mu_n + \frac{1}{3}\mu_e, \label{eq:CS1}\\
\mu_p +\frac{ 2\varepsilon_{\rm{Coul}} }{\left(1 - \chi\right) \, \delta n_c}
  &=& \mu_n - \mu_e, \label{eq:CN1}\\
\mu_\mu &=& \mu_e. \label{eq:CE1}
\end{eqnarray}
Minimizing $w$ with respect to the volume fraction $\chi$ leads to the equilibrium condition
for the pressures,
\begin{eqnarray}
P_{\rm{HP}} &=& P_{\rm{QP}} -\frac{2\varepsilon_{\rm{Coul}}}{\delta n_c}
  \left[ \frac{1}{3\chi}\left( 2n_u - n_d - n_s \right)+\frac{1}{1-\chi}n_p\right] \notag \\
& &  \mp \frac{\varepsilon_{\rm{Coul}} }{\chi_{\rm{in}}}
  \left(3+\chi_{\rm{in}}\frac{{\Phi}^{^{\prime }}}{\Phi}\right),
\label{eq:CP1}
\end{eqnarray}
where the sign of the last term is \textquotedblleft $-$\textquotedblright\ for droplet,
rod, and slab configurations, while it is\textquotedblleft $+$\textquotedblright\ for tube
and bubble configurations.
It is clear that equilibrium conditions for two-phase coexistence are
altered due to the inclusion of surface and Coulomb terms in the minimization
procedure, and, as a result, they are different from the Gibbs equilibrium conditions.
Some additional terms appearing in the equilibrium equations are caused by
the surface and Coulomb contributions.
If we neglect the finite-size effects by taking the limit $\sigma \rightarrow 0$,
these additional terms vanish and the equilibrium equations reduce to the
Gibbs conditions.
Generally, the pressure of the mixed phase can be calculated from the thermodynamic relation,
$P_{\rm{MP}} = n_b^{2}\frac{\partial \left(\varepsilon_{\rm{MP}}/n_b\right) }{\partial n_b}$,
which is somewhat different from $P_{\rm{HP}}$ and $P_{\rm{QP}}$.
This is similar to the case of nuclear liquid-gas phase transition
at subnuclear densities~\cite{Bao14b,Baym71,Latt85,Latt91}.
Furthermore, minimizing $w$ with respect to the size $r_D$ yields the well-known
relation ${\varepsilon}_{\rm{surf}}=2{\varepsilon}_{\rm{Coul}}$, which leads to
the formula for the size of the inner phase,
\begin{eqnarray}
\label{eq:rd}
r_D &=& \left[\frac{\sigma{D}}{e^2\left(\delta n_c\right)^{2}\Phi}\right]^{1/3}.
\end{eqnarray}

The properties of hadron-quark pasta phases are obtained by solving the above
equilibrium equations at a given baryon density $n_b$.
We compare the energy density of the mixed phase with different pasta configurations
and then determine the most stable configuration with the lowest energy density.
The hadron-quark mixed phase exists only in the density range where its energy density
is lower than that of both hadronic matter and quark matter.

\section{Results and discussion}
\label{sec:5}

In this section, we present the numerical results of the hadron-quark pasta phases
using the EM method described in the previous section. To examine the dependence
of results on the bag constant of quark matter, we calculate and compare
the results with $B^{1/4}=210.854$~MeV and $B^{1/4}=180$~MeV in the present work.
Note that $B^{1/4}=210.854$~MeV is determined in the QMC model
by fitting the free nucleon properties. As for the surface tension at the hadron-quark
interface, we self-consistently calculate its value by using the MRE method within the
MIT bag model. Since the quark degrees of freedom are explicitly involved
in the QMC model, the descriptions of hadronic and quark phases are considered to be
consistent with each other at the quark level.
The properties of neutron stars are then calculated using the EOS with the inclusion of
quarks at high densities.

\begin{figure*}[htbp]
\includegraphics[clip,width=15.6 cm]{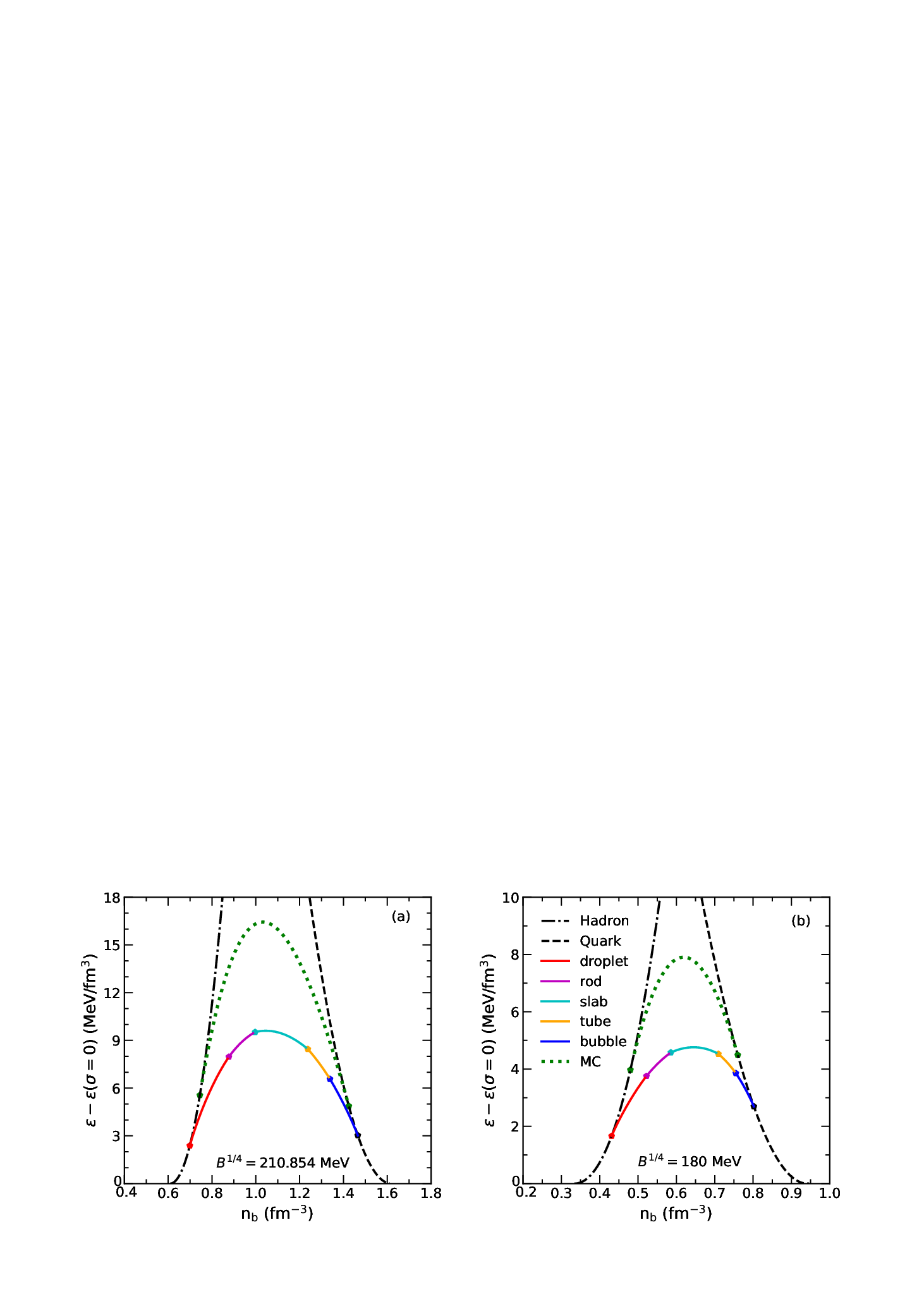}
\caption{Energy densities of the mixed phase obtained using the
EM method relative to those of the Gibbs construction ($\sigma=0$).
The filled circles indicate the transition between different pasta phases.
The results of the Maxwell construction (MC) are shown by green dotted lines.}
\label{fig:1enb}
\end{figure*}

\subsection{Hadron-quark pasta phases}
\label{sec:5-1}

Due to the competition between the surface and Coulomb energies, the geometric structure of
a hadron-quark mixed phase is expected to change from droplet to rod, slab, tube, and
bubble with increasing baryon density. In the present calculations, we employ the EM method
to investigate the pasta phases, where the equilibrium conditions between coexisting
phases are derived by the minimization of the total energy including surface and Coulomb
contributions. Since the surface and Coulomb energies are positive, the energy densities
of pasta phases are higher than that of the Gibbs construction without finite-size effects.
Generally, the energy density difference between different pasta shapes is rather small
compared to the total energy density, but it is crucial to determine the transition of
pasta shapes.
In Fig.~\ref{fig:1enb}, we show the energy densities of pasta phases obtained using
the EM method relative to those of the Gibbs construction (i.e., $\sigma = 0$).
The filled circles indicate the transition between different pasta phases.
For comparison, the energy densities of pure hadronic and pure quark phases
are respectively plotted by black dot-dashed and dashed lines, whereas
the results of the Maxwell construction are shown by green dotted lines.
To examine the influence of the bag constant in quark matter, we present
the results with $B^{1/4}=210.854$~MeV and $B^{1/4}=180$~MeV in the left and right
panels, respectively. We note that the onset of the mixed phase occurs when its
energy density becomes lower than that of pure hadronic matter.
In the case of $B^{1/4}=210.854$~MeV that is consistently used in the QMC model
and quark matter, the formation of quark droplets occurs at $0.7\, \rm{fm}^{-3}$,
which is significantly higher the onset of the mixed phase at $0.6\, \rm{fm}^{-3}$
obtained in the Gibbs construction. This is because the finite-size effects like surface
and Coulomb contributions increase the energy density and delay the appearance of the mixed
phase.  As $n_{b}$ increases, other pasta shapes, such as rod, slab, tube, and bubble, may
appear when one has the lowest energy density among all configurations.
Furthermore, the mixed phase ends at $1.47\, \rm{fm}^{-3}$
where the energy density of pure quark matter becomes lower than that of pasta phases.
In the right panel of Fig.~\ref{fig:1enb}, we can see that using a smaller bag constant
of quark matter $B^{1/4}=180$~MeV leads to an early onset of the structured mixed
phase at $0.43\, \rm{fm}^{-3}$ and pure quark phase at
$0.8\, \rm{fm}^{-3}$, while the behavior is qualitatively similar to
that displayed in the left panel of Fig.~\ref{fig:1enb}.
The influence of the bag constant $B$ on the hadron-quark phase transition
has been extensively discussed in the literature~\cite{Sche00,Baym18}.
It is seen that the energy density in the Maxwell construction is clearly higher than
that of pasta phases, so the Maxwell construction is disfavored in the present calculations.

\begin{figure}[htbp]
\includegraphics[clip,width=8.6 cm]{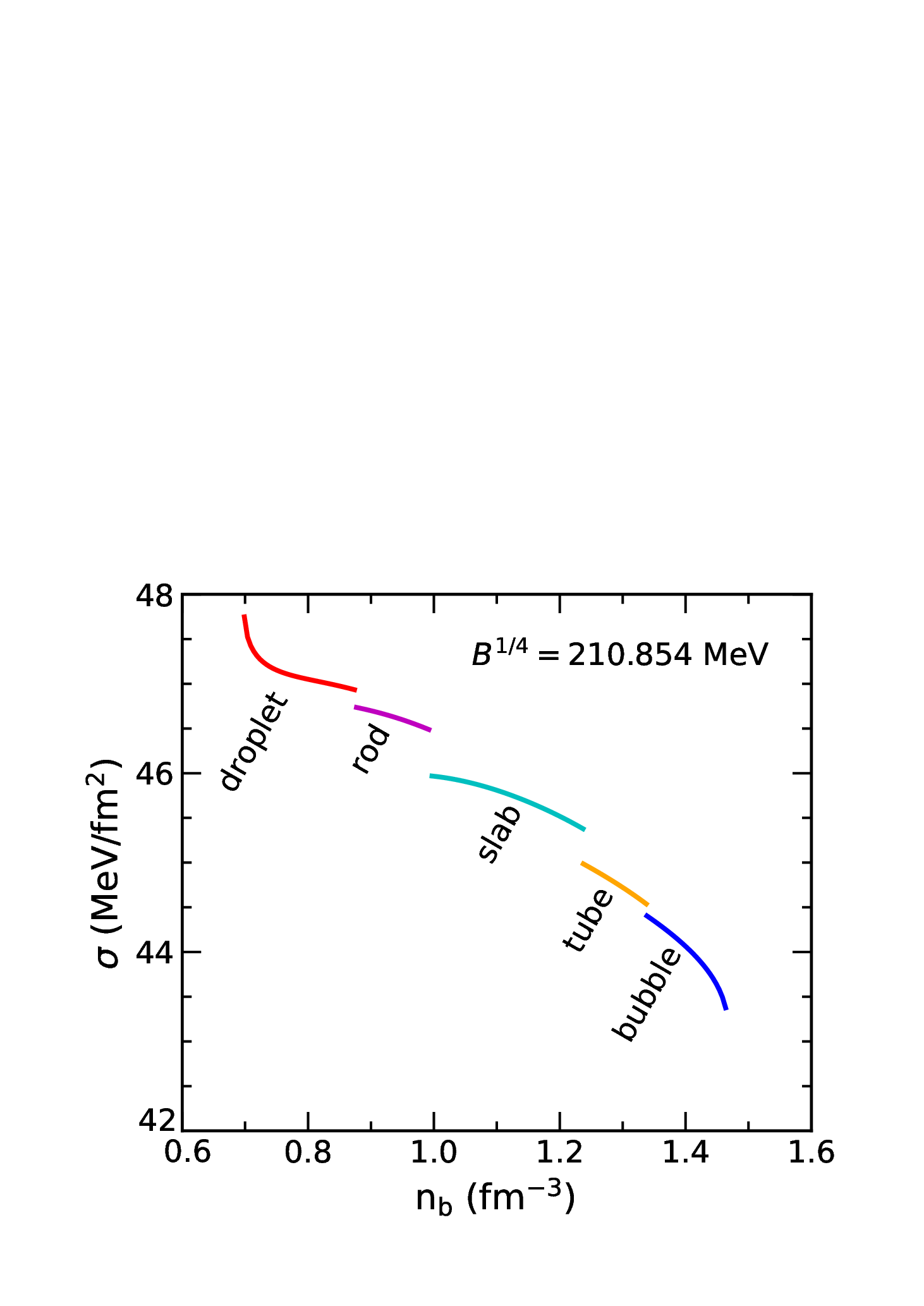}
\caption{Surface tension $\sigma$ as a function of the baryon
density $n_{b}$ obtained in pasta phases. }
\label{fig:2signb}
\end{figure}

\begin{figure}[htbp]
\includegraphics[clip,width=8.6 cm]{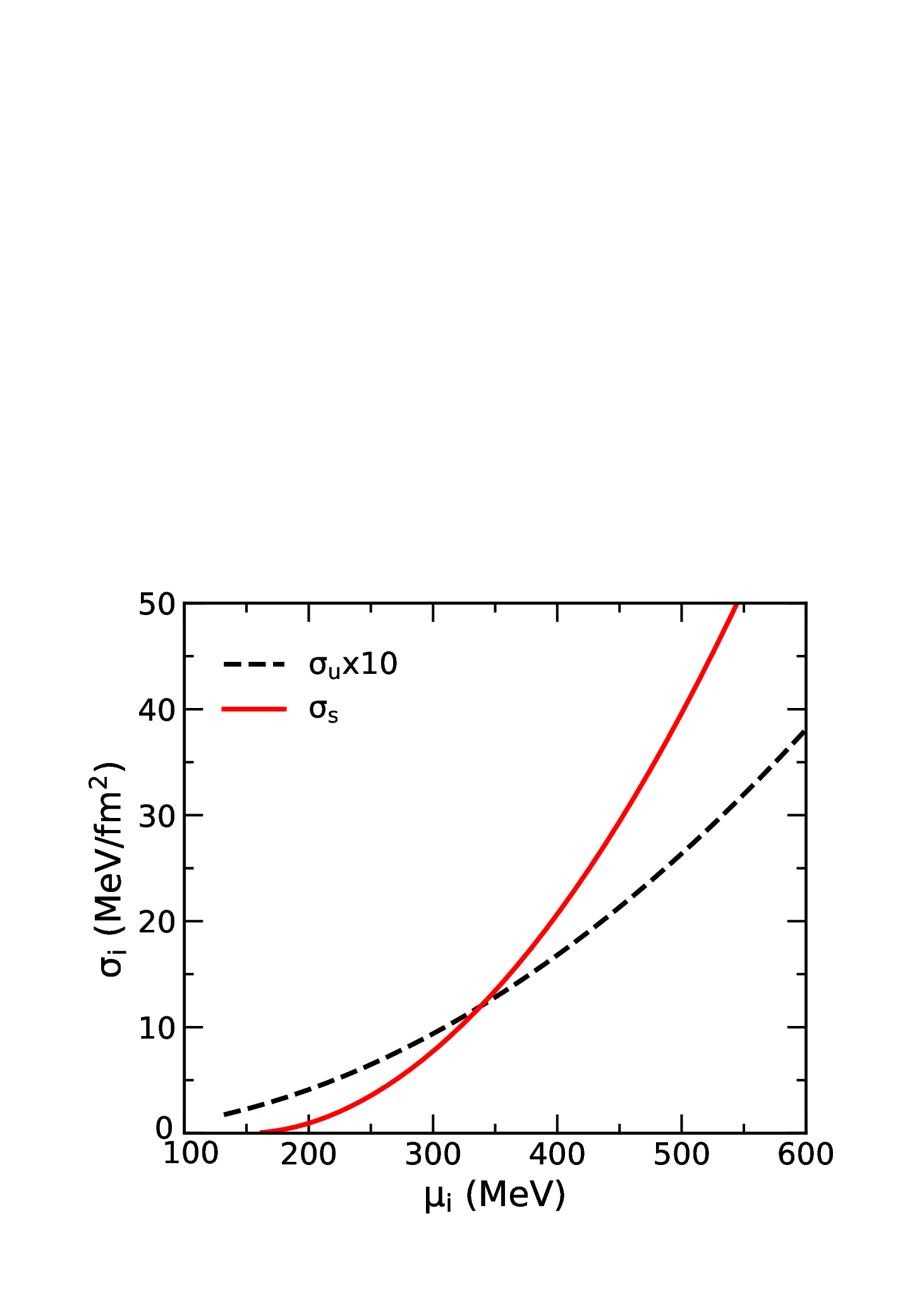}
\caption{Surface tension of the quark flavor $i$ as a function of its
chemical potential calculated by Eq.~(\ref{eq:sigma}). }
\label{fig:3sigmu}
\end{figure}

\begin{figure}[htbp]
\includegraphics[clip,width=8.6 cm]{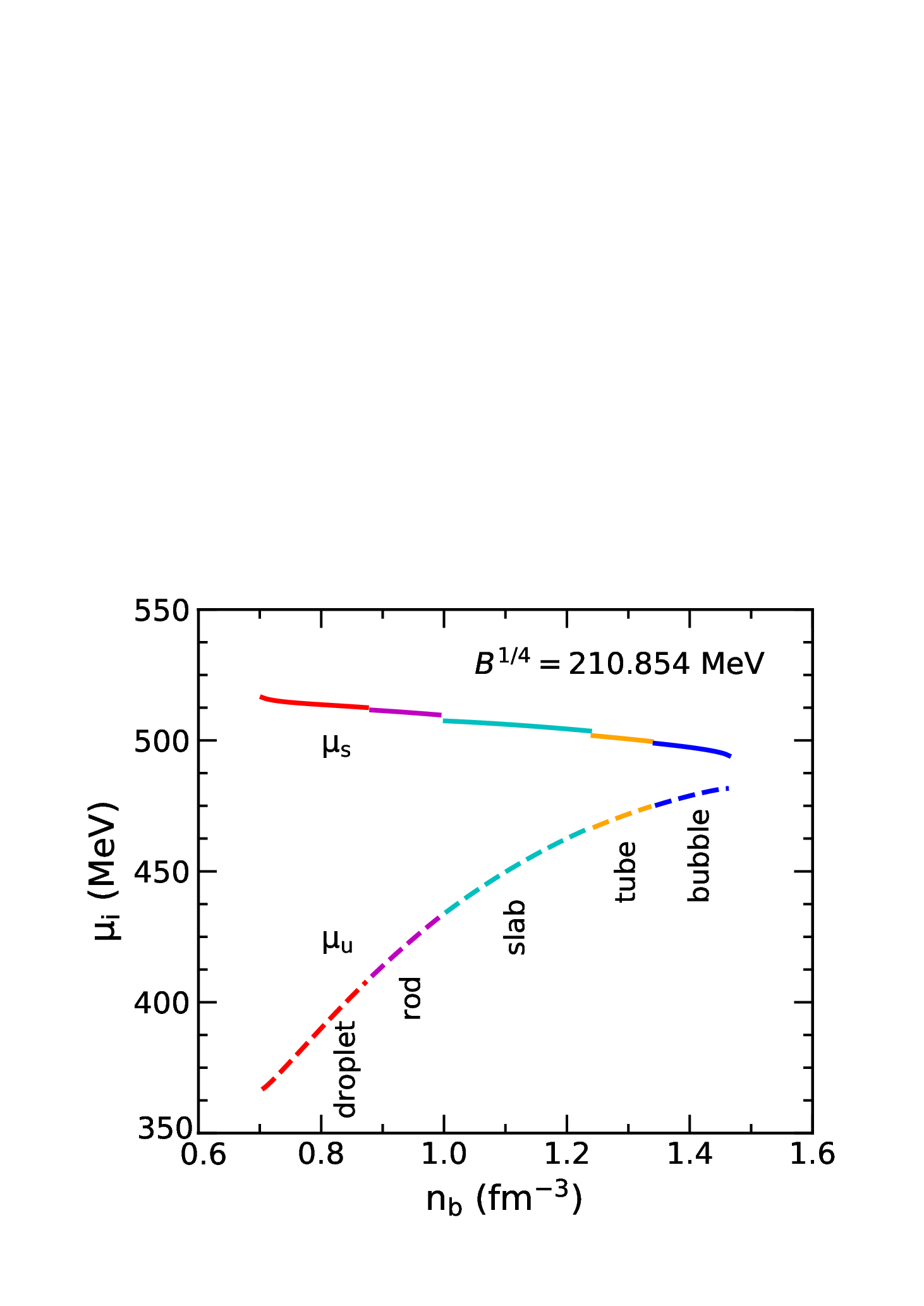}
\caption{Chemical potential of the quark flavor $i$ as a function of the baryon
density $n_{b}$ obtained in pasta phases. }
\label{fig:4munb}
\end{figure}

\begin{figure}[htbp]
\includegraphics[clip,width=8.6 cm]{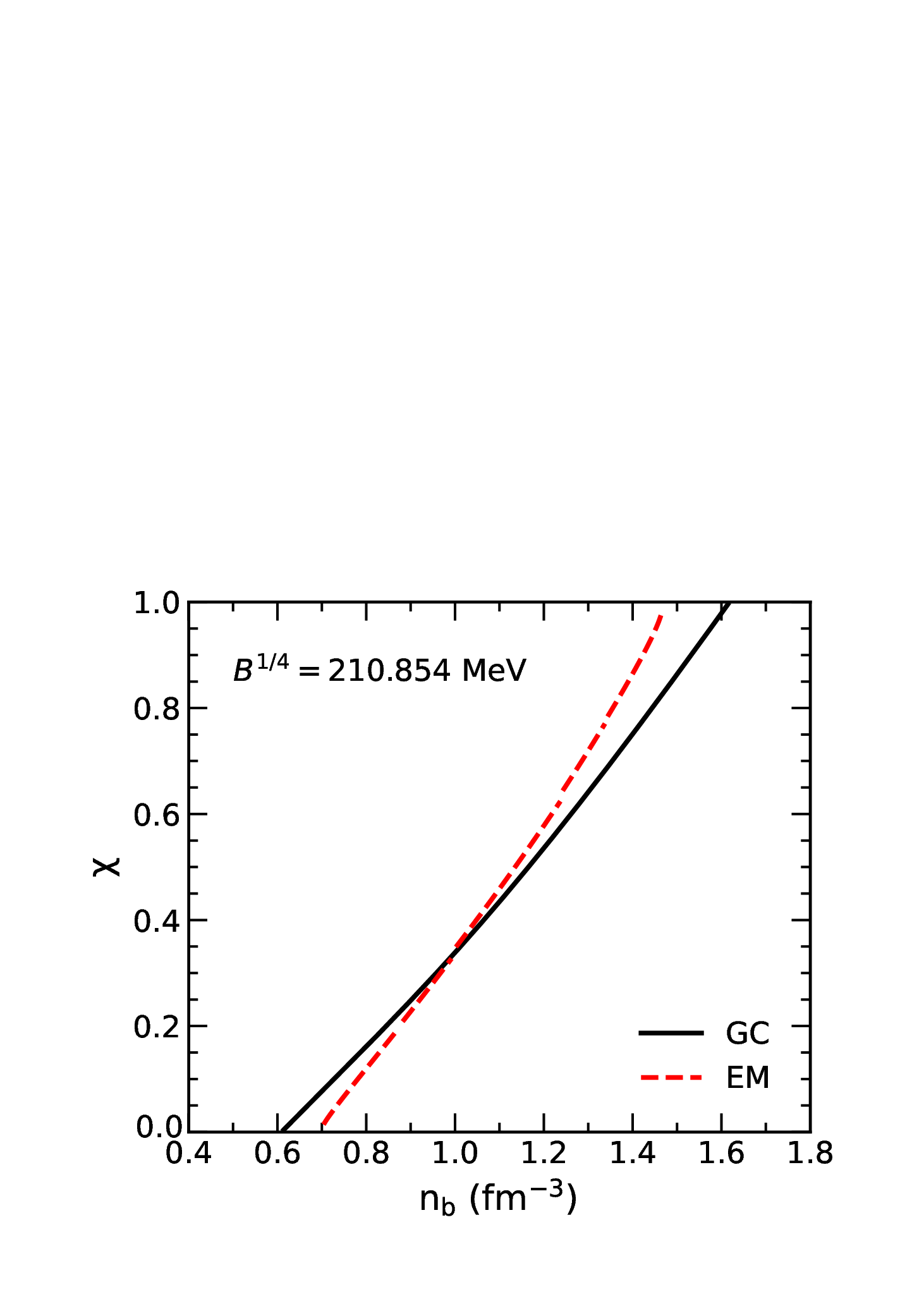}
\caption{Volume fraction of the quark phase $\chi$ as a function of the baryon
density $n_{b}$ obtained in the mixed phase.
The results using the EM method are compared to those of the Gibbs construction (GC). }
\label{fig:5unb}
\end{figure}

\begin{figure}[htbp]
\includegraphics[clip,width=8.6 cm]{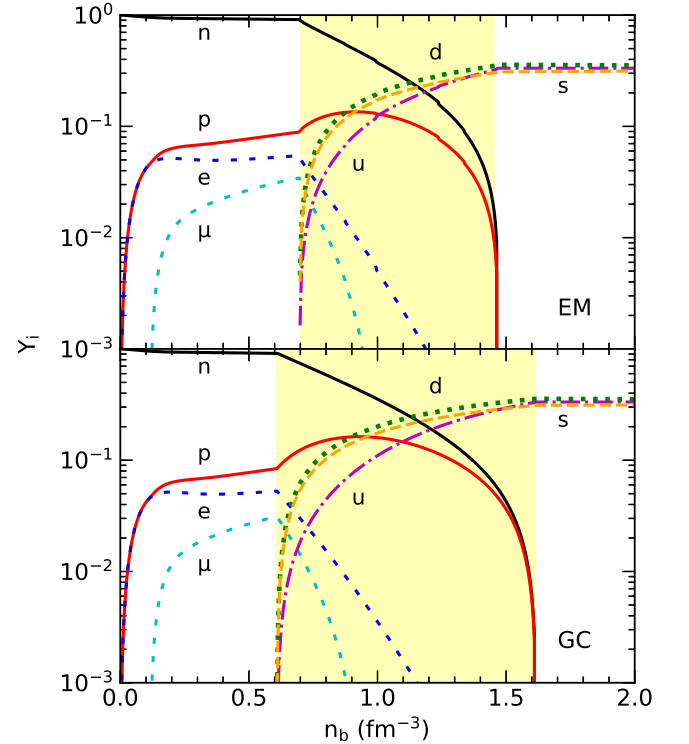}
\caption{Particle fraction $Y_i$ as a function of the baryon density $n_{b}$
obtained using the EM method (upper panel) and Gibbs construction (lower panel).
The shaded areas denote the mixed phase regions. }
\label{fig:6yinb}
\end{figure}


The surface tension at the hadron-quark interface plays a crucial role in
determining the structure of pasta phases. In Fig.~\ref{fig:2signb}, we display
the surface tension $\sigma$ as a function of the baryon density $n_{b}$
obtained in pasta phases with $B^{1/4}=210.854$~MeV.
In the present work, we use the MRE method to calculate the surface tension
that is the sum over quark flavors,
$\sigma=\sum_{i=u,d,s}\sigma_i$, with $\sigma_i$ given by Eq.~(\ref{eq:sigma}).
One can see that $\sigma$ slightly decreases with increasing $n_{b}$ and
its value lies in the range of $43-48$~MeV/fm$^{2}$.
This behavior can be understood by analyzing Eq.~(\ref{eq:sigma}),
where $\sigma_i$ is a function of the quark mass $m_i$ and chemical potential $\mu_i$.
The dominant contribution to the surface tension comes from the $s$ quark,
since its mass is much larger than that of $u$ and $d$ quarks.
In Fig.~\ref{fig:3sigmu}, we plot $\sigma_i$ as a function of $\mu_i$ for $i=s$ and $i=u$.
It is found that $\sigma_s$ is about one order higher than $\sigma_u$,
and it increases significantly as the chemical potential increases.
However, in the pasta phases, $\mu_s$ shown in Fig.~\ref{fig:4munb} decreases
with increasing $n_{b}$, which leads to the decline of the surface tension $\sigma$
as shown in Fig.~\ref{fig:2signb}.

In Fig.~\ref{fig:5unb}, we show the volume fraction of the quark phase $\chi$ as a function
of the baryon density $n_{b}$ in the mixed phase with $B^{1/4}=210.854$~MeV.
The results obtained in the EM method are compared to those of the Gibbs construction.
It is shown that $\chi$ continuously increases with increasing $n_{b}$ in the mixed phase.
The behavior of $\chi$ in the EM method is very similar to that of the Gibbs construction,
but the density range of pasta phases is reduced due to the inclusion of finite-size
effects. There is no large jump in $\chi$ at the transition point of pasta configurations,
so the hadron-quark phase transition described in the EM method is relatively smooth.
Because of the monotonic increase of $\chi$, more and more hadronic matter is
converted into quark matter during the phase transition.
In Fig.~\ref{fig:6yinb}, we display the relative particle fractions $Y_i=n_i/n_b$
as a function of the baryon density $n_{b}$ with $B^{1/4}=210.854$~MeV.
The results obtained in the EM method and Gibbs construction are shown in the upper
and lower panels, respectively. The shaded areas denote the mixed phase regions.
One can see that the matter at low densities consists
of neutrons, protons, and electrons, whereas the muons appear at $n_{b}=0.11\, \rm{fm}^{-3}$
playing the same role as electrons. When the quark matter is present in the mixed phase,
the fractions of quarks, $Y_u$, $Y_d$, and $Y_s$, increase rapidly together with a decrease of
neutron fraction $Y_n$. Meanwhile, $Y_e$ and $Y_\mu$ decrease significantly, since
the quark matter is negatively charged that can take the role of electrons
to satisfy the constraint of global charge neutrality.
On the other hand, the hadronic matter in the mixed phase is positively charged,
so the proton fraction $Y_p$ increases at the beginning of the mixed phase.
As $n_{b}$ increases, $Y_p$ and $Y_n$ decrease to very low values due to
increasing $\chi$ as shown in Fig.~\ref{fig:5unb}.
At sufficiently high densities, the hadronic matter completely disappears and the
transition to a pure quark phase occurs.
It is seen that $Y_u \approx Y_d \approx Y_s \approx 1/3$ is achieved in the pure
quark phase, which is a result of the chemical equilibrium and charge neutrality
given in Eqs.~(\ref{eq:qchq}) and (\ref{eq:qnc}).

\begin{figure}[htb]
\includegraphics[clip,width=8.6 cm]{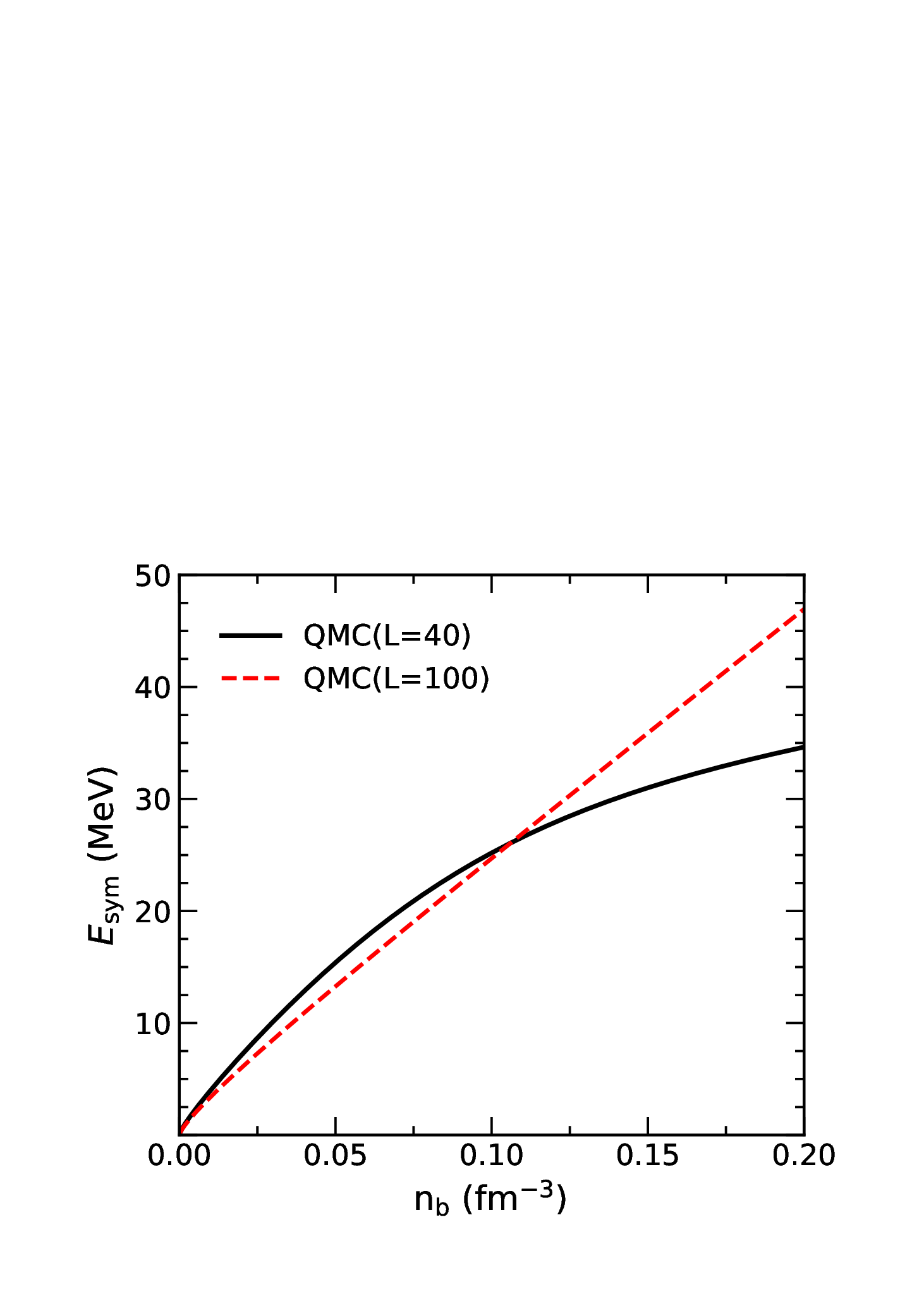}
\caption{Symmetry energy $E_{\text{sym}}$ as a function of
the baryon density $n_b$ obtained in the QMC($L$=40) and QMC($L$=100) models.}
\label{fig:7Esym}
\end{figure}

\begin{figure}[htbp]
\includegraphics[clip,width=8.6 cm]{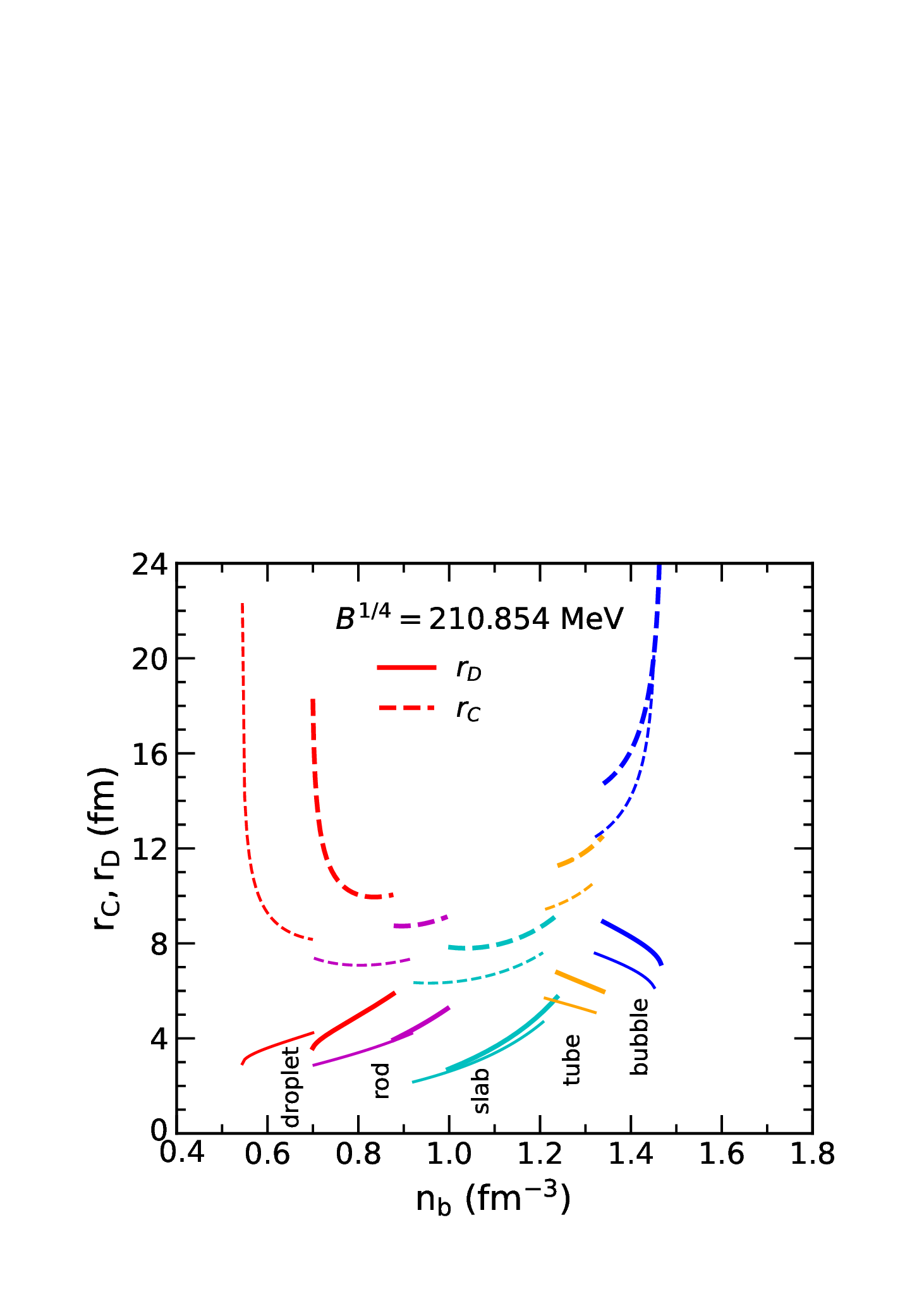}
\caption{Size of the Wigner-Seitz cell ($r_{C}$) and that of the inner
phase ($r_{D}$) as a function of $n_{b}$ obtained using the EM method.
The results with QMC($L$=40) and QMC($L$=100) are shown by thick and thin lines,
respectively. }
\label{fig:8rnb}
\end{figure}

\subsection{Nuclear symmetry energy effects}
\label{sec:5-2}

It is well known that the nuclear symmetry energy $E_{\textrm{sym}}$ and its
slope $L$ can significantly affect the properties of neutron stars,
especially the neutron-star radius, tidal deformability, and crust structure,
which are particularly sensitive to the slope parameter $L$~\cite{Fatt18,Bao15,Oert17}.
Many efforts have been devoted to constraining the values of $E_{\rm sym}$ and $L$
at saturation density based on astrophysical observations and terrestrial nuclear
experiments (see Refs.~\citep{Oert17,LiBA08,Tews17} and references therein).
According to available constraints summarized in Ref.~\cite{Oert17},
it was found that the most probable values for the symmetry energy and its
slope at saturation density are $E_{\rm sym}=31.7\pm 3.2$~MeV
and $L=58.7\pm 28.1$~MeV, respectively,
with a much larger error for $L$ than that for $E_{\rm sym}$.
In order to explore the influence of nuclear symmetry energy on the hadron-quark
pasta phases, we adjust the isovector couplings, $g_{\rho}$ and ${\Lambda}_{\rm{v}}$,
so as to obtain a stiff EOS in the QMC model and compare with the soft one used.
By taking ${\Lambda}_{\rm{v}}=0$ and $g_{\rho}=4.7086$, we obtain the QMC model
with $E_{\rm{sym}}=35.9$~MeV and $L=100$~MeV at saturation density,
whereas the isoscalar saturation properties remain unchanged.
In Fig.~\ref{fig:7Esym}, we show the symmetry energy $E_{\rm{sym}}$
as a function of the baryon density $n_b$ obtained in the QMC($L$=40)
and QMC($L$=100) models. It is found that the two models have the same
$E_{\text{sym}}$ at $n_b\simeq 0.106\, \rm{fm}^{-3}$, but they display
rather different density dependence due to different slope parameter $L$.
It is noteworthy that the binding energies of finite
nuclei are directly related to the symmetry energy at subsaturation
density of $0.10-0.11\, \rm{fm}^{-3}$~\cite{Bao14b,Zhan13},
which corresponds to the average density in nuclei.
Therefore, the QMC($L$=40) and QMC($L$=100) models are expected to provide
similar descriptions for finite nuclei, since they have the same isoscalar
properties and equal values of $E_{\text{sym}}$ at $n_b\simeq 0.106\, \rm{fm}^{-3}$.
On the other hand, the different slope parameter $L$ can significantly
alter the stiffness of neutron-star matter EOS at high densities,
which may lead to considerable differences in the hadron-quark
pasta phases inside neutron stars.

We calculate and compare the properties of the hadron-quark pasta phases
by using the QMC($L$=40) and QMC($L$=100) models for the hadronic matter.
In Fig.~\ref{fig:8rnb}, we display the size of the Wigner-Seitz cell ($r_C$) and
that of the inner phase ($r_D$) as a function of the baryon density $n_{b}$
obtained using the EM method with $B^{1/4}=210.854$~MeV.
The results with QMC($L$=40) and QMC($L$=100) are shown by thick and thin lines,
respectively. It has been reported in Ref.~\citep{Wu19} that a larger
symmetry energy slope $L$ in hadronic matter corresponds to an earlier onset of
the hadron-quark phase transition. In the present work, a similar trend is
seen, namely the formation of quark droplets with QMC($L$=100)
occurs at a lower density compared to the case with QMC($L$=40).
Furthermore, there are also differences between QMC($L$=40) and QMC($L$=100)
in the size of the pasta structure, especially in $r_C$.
One can see that as the density increases, $r_D$ in the droplet, rod, and slab
phases increases, whereas $r_D$ in the tube and bubble phases decreases.
This is consistent with the monotonic increase of $\chi$ shown in Fig.~\ref{fig:5unb}.
There are obvious discontinuities in $r_D$ and $r_C$ at the transition between
different pasta shapes, which exhibit the character of the first-order transition.
Generally, a large surface tension $\sigma$ leads to a large radius of quark droplets.
In the present work, we use the MRE method to calculate $\sigma$ that is in the
range of $43-48$~MeV/fm$^{2}$, which yields the radius of quark droplets as
$r_D \approx 3-6 \, \rm{fm}$.

To make a detailed comparison between QMC($L$=40) and QMC($L$=100),
we present in Table~\ref{tab:1} the onset densities of the hadron-quark pasta
phases and pure quark matter obtained using the EM method.
It is found that the onset densities with QMC($L$=100) are noticeably smaller than
those with QMC($L$=40), and the differences gradually decrease from the droplet
to pure quark phases. This is because the fraction of hadronic matter monotonically
decreases during the hadron-quark phase transition, so that the influence of nuclear
symmetry energy becomes weaker and weaker.
Due to the same reason, the influence of the bag constant $B$ gets stronger and stronger
during the phase transition. One can see that the onset densities of pure quark matter
with $B^{1/4}=180$~MeV are much smaller than those with $B^{1/4}=210.854$~MeV.
We find that the results are very sensitive to the bag constant adopted.
By using a small bag constant of $B^{1/4}=180$~MeV, the hadron-quark pasta phases
are significantly shifted toward the lower density region compared to the results
with $B^{1/4}=210.854$~MeV.

\begin{table*}[htb]
\caption{Onset densities of the hadron-quark pasta phases and pure quark
matter obtained from different models for describing the hadronic and quark phases.}
\begin{center}
\begin{tabular}{llcccccc}
\hline\hline
QMC model & Bag model & \multicolumn{6}{c}{Onset density (fm$^{-3}$)}      \\
\cline{3-8}
      &    &droplet &rod    &slab   &tube   &bubble & quark    \\
\hline
$L=40 $ MeV  & $B^{1/4}=210.854$ MeV &0.700  &0.830 &0.997 &1.238 &1.339 &1.466  \\
$L=100$ MeV  & $B^{1/4}=210.854$ MeV &0.545  &0.702 &0.921 &1.211 &1.321 &1.453  \\
$L=40 $ MeV  & $B^{1/4}=180$ MeV     &0.431  &0.522 &0.586 &0.710 &0.755 &0.803  \\
$L=100$ MeV  & $B^{1/4}=180$ MeV     &0.306  &0.428 &0.514 &0.681 &0.737 &0.791  \\
\hline\hline
\end{tabular}
\label{tab:1}
\end{center}
\end{table*}

\subsection{Properties of neutron stars}
\label{sec:5-3}

\begin{figure*}[htbp]
\includegraphics[clip,width=15.6 cm]{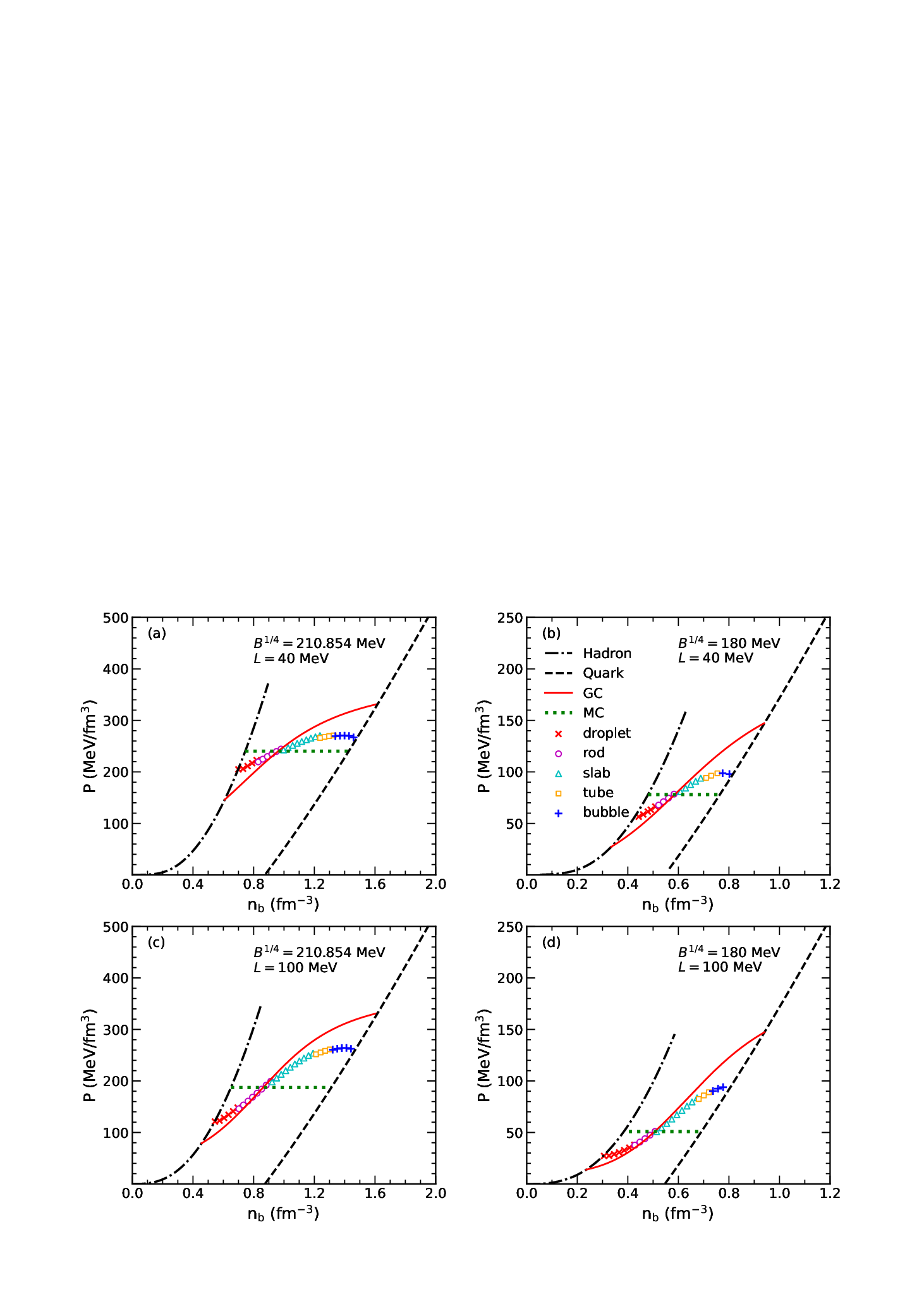}
\caption{Pressures as a function of the baryon density $n_{b}$
for hadronic, mixed, and quark phases.
The results of pasta phases obtained using the EM method are compared to
those of the Gibbs and Maxwell constructions. }
\label{fig:9pnb}
\end{figure*}

\begin{figure*}[htbp]
\includegraphics[clip,width=15.6 cm]{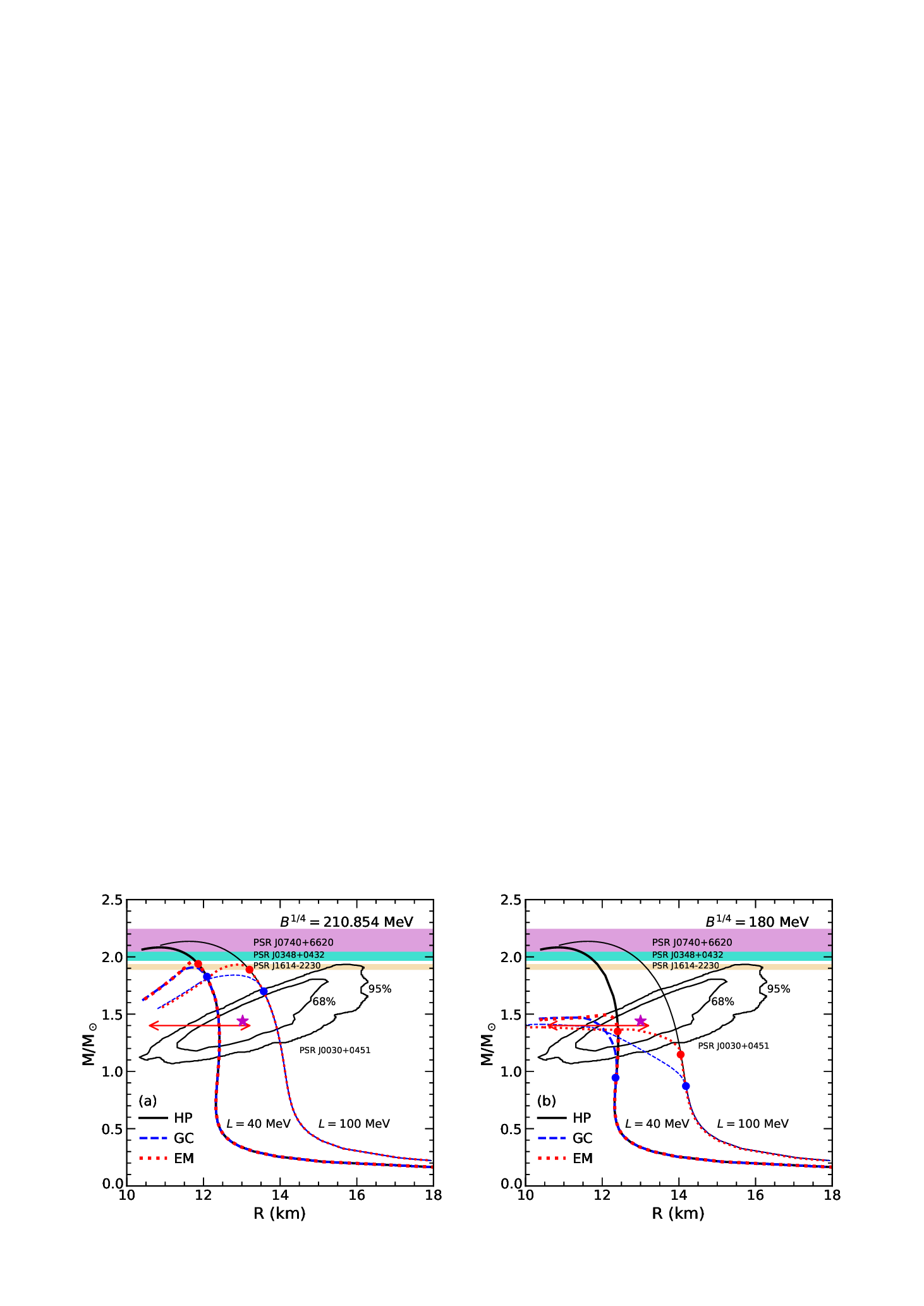}
\caption{Mass-radius relations of neutron stars for different EOS.
The results of a pure hadronic phase (solid lines) are compared to those including
quarks in the EM method (dotted lines) and Gibbs construction (dashed lines)
with $B^{1/4}=210.854$~MeV (left panel) and $B^{1/4}=180$~MeV (right panel).
The filled circles indicate the onset of the star containing a hadron-quark mixed phase.
The results with QMC($L$=40) and QMC($L$=100) are shown by thick and thin lines, respectively.
The horizontal bars indicate the observational constraints of
PSR J1614--2230~\cite{Demo10,Fons16,Arzo18}, PSR J0348+0432~\cite{Anto13},
and PSR J0740+6620~\cite{Crom19}.
The simultaneous measurement of the mass and radius for PSR J0030+0451 by {\it NICER}
is also shown by the star with 68\% and 95\% confidence intervals~\citep{Mill19}.
The horizontal line with arrows at both ends represents the constraints on $R_{1.4}$
inferred from GW170817~\cite{Abbo18}.
}
\label{fig:10mr}
\end{figure*}

\begin{figure}[htbp]
\includegraphics[clip,width=8.6 cm]{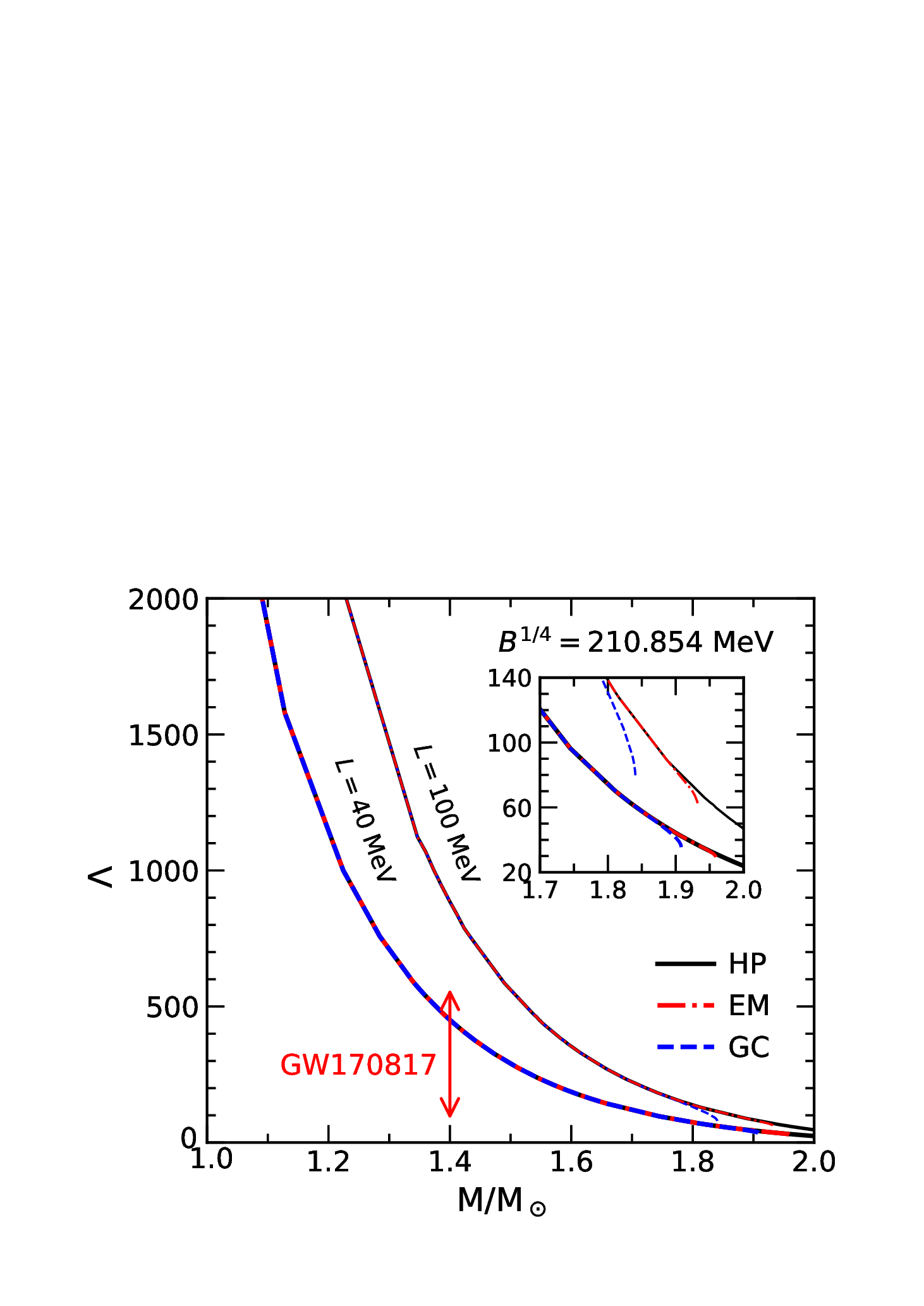}
\caption{Dimensionless tidal deformability $\Lambda$
as a function of the neutron-star mass $M$.
The vertical line with arrows at both ends represents the constraints on $\Lambda_{1.4}$
from the analysis of GW170817~\cite{Abbo18}. }
\label{fig:11lm}
\end{figure}

To investigate the influence of quark matter on neutron-star properties,
we provide the EOS including the hadron-quark pasta phases and pure quark
matter at high densities.
In Fig.~\ref{fig:9pnb}, we plot the pressures as a function of the baryon
density $n_{b}$ for hadronic, mixed, and quark phases.
The results with $B^{1/4}=210.854$~MeV and $B^{1/4}=180$~MeV are displayed in
the left and right panels, respectively.

The results with QMC($L$=40) shown in the upper panels are compared to those
with QMC($L$=100) in the lower panels.

The pressures of pasta phases obtained using the EM method are compared to
those of the Gibbs and Maxwell constructions. It can be seen that the pressures
of pasta phases lie between those of the Gibbs and Maxwell constructions.
The pressures with the Maxwell construction remain constant during
the phase transition, whereas those with the Gibbs construction increase
with $n_{b}$ over a relatively broad range.
The influence of the bag constant $B$ on the EOS can be observed by comparing
the left and right panels. It is clearly shown that a smaller $B$ leads to
early onset of the mixed phase and pure quark phase, namely the mixed phase
is shifted toward a lower density region. As a result, the pressures of pasta
phases with $B^{1/4}=180$~MeV are much lower than those with $B^{1/4}=210.854$~MeV.

Meanwhile, the effect of nuclear symmetry energy on the EOS is seen by comparing
the upper and lower panels. In the case with the QMC($L$=100) model (lower panels),
the onset of the mixed phase occurs at lower densities than with the QMC($L$=40)
model (upper panels), but there is no visible difference at the end of the
mixed phase.

One can see that the behavior of $P$ in pasta phases is similar to that of the Gibbs
construction, and no abrupt jump is observed at the transition between different
pasta configurations.

The properties of neutron stars, such as the mass-radius relations and
tidal deformabilities, can be obtained by solving the Tolman-Oppenheimer-Volkoff (TOV)
equation using the EOS over a wide range of densities.
In the present work, we adopt the Baym--Pethick--Sutherland (BPS) EOS~\cite{BPS71}
for the outer crust below the neutron drip density, while the inner crust EOS is based
on a Thomas--Fermi calculation using the TM1($L$=40) effective interaction
of the RMF model~\cite{Ji19}. The crust EOS is matched to the QMC EOS of uniform neutron-star
matter at the crossing point of the two segments.
At high densities, the hadron-quark pasta phases are taken into account by using
the QMC model for the hadronic phase and the MIT bag model for the quark phase.
In Fig.~\ref{fig:10mr}, we present the resulting mass-radius relations of neutron stars
with the inclusion of quarks at high densities.
The results with $B^{1/4}=210.854$~MeV and $B^{1/4}=180$~MeV are shown in
the left and right panels, respectively.
The mass measurements of
PSR J1614--2230 ($1.908 \pm 0.016  M_\odot$)~\cite{Demo10,Fons16,Arzo18},
PSR J0348+0432 ($2.01  \pm 0.04   M_\odot$)~\cite{Anto13}, and
PSR J0740+6620 ($2.14^{+0.10}_{-0.09}  M_\odot$)~\cite{Crom19}
are indicated by the horizontal bars.
The simultaneous measurement of the mass and radius for PSR J0030+0451 by {\it NICER}
is also shown by the star with 68\% and 95\% confidence intervals~\citep{Mill19}.

The constraints on $R_{1.4}$ inferred from GW170817~\cite{Abbo18} are indicated
by the horizontal line with arrows at both ends.

We find that the maximum mass of neutron stars using a pure hadronic EOS in
the QMC($L$=40) model is about $2.08 M_\odot$, which is considerably reduced
as the hadron-quark phase transition is included.
The star masses obtained in the EM method are slightly larger
than those of the Gibbs construction due to finite-size effects.
By comparing the left and right panels, it is clear that the results are sensitive to
the bag constant $B$ adopted.
In the case of $B^{1/4}=210.854$~MeV, the EM method with QMC($L$=40)
leads to a maximum mass of $1.96  M_\odot$,
where the hadron-quark pasta phases could be formed in the interior of massive
stars with $M > 1.94 M_\odot$, but no pure quark phase exists.
In fact, a canonical $1.4 M_\odot$ neutron star would not be affected by the quark phase,
since its central density is lower than the onset of a hadron-quark mixed phase.
However, in the case of $B^{1/4}=180$~MeV, the maximum mass obtained in the EM method
with QMC($L$=40) is about $1.49  M_\odot$, which is much lower than
the constraint of $2 M_\odot$.
Therefore, the early onset of the mixed phase with $B^{1/4}=180$~MeV
is disfavored since it leads to a large reduction of the maximum neutron-star mass.

It is interesting to see the effect of nuclear symmetry energy on the neutron-star
properties. When the QMC($L$=100) model is adopted, the maximum mass of neutron stars
is slightly higher than that in the QMC($L$=40) model, but the radius is
significantly increased. This is because the QMC($L$=100) model provides a stiff EOS,
which results in relatively large radii of neutron stars. It is shown that the large
radius of $R_{1.4}= 13.9$ km obtained in the QMC($L$=100) model is
disfavored by the constraints from GW170817.
In contrast, we obtain a radius of $R_{1.4}= 12.43$ km in the QMC($L$=40) model,
which is compatible with recent observational constraints from {\it NICER} and GW170817.

The dimensionless tidal deformability of a neutron star can be calculated from
\begin{eqnarray}
\label{eq:td}
\Lambda=\frac{2}{3}k_2 \left(R/M\right)^{5},
\end{eqnarray}
where $k_2$ is the tidal Love number which is computed together with the TOV equation
as described in Refs.~\cite{Ji19,Hind08,Post10}.
In Fig.~\ref{fig:11lm}, we display the dimensionless tidal deformability $\Lambda$
as a function of the gravitational mass $M$ of the star.
It is shown that $\Lambda$ decreases rapidly with increasing $M$.

The results with the inclusion of quarks using $B^{1/4}=210.854$~MeV are compared
to those using a pure hadronic EOS in the QMC($L$=40) and QMC($L$=100) models.
It is seen that the effect of the hadron-quark phase transition is almost invisible,
since the mixed phase is present only in massive stars.
For a canonical $1.4 M_\odot$ neutron star, we obtain
$\Lambda_{1.4} = 450$ in the QMC($L$=40) model, which is consistent
with the constraints from the analysis of GW170817~\cite{Abbo18},
while $\Lambda_{1.4} = 885$ obtained in the QMC($L$=100) model is incompatible
with the constraints. One can see that a small discrepancy in $\Lambda$ using
the EM method is observed, and it becomes more obvious in the Gibbs
construction. The reduction of $\Lambda$ is a combined result
of changes in the Love number $k_2$ and compactness parameter $M/R$ caused
by the appearance of the hadron-quark phase transition.

\section{Conclusions}
\label{sec:6}

In the present work, we have studied the properties of the hadron-quark pasta
phases, which are expected to occur in the interior of massive neutron stars.
We have employed the QMC model to describe the hadronic phase, where the internal
quark structure of the nucleon is explicitly taken into account based on the MIT
bag model. For the description of the quark phase, we have adopted the same bag model
as the one used in the QMC framework, so that the coexisting hadronic and quark
phases are treated in a consistent way.
We have used the Wigner--Seitz approximation to describe the hadron-quark pasta phases,
where the system is divided into equivalent cells with a given geometric symmetry.
The hadronic and quark phases inside the cell are assumed to have constant densities
and are separated by a sharp interface. We computed the surface tension $\sigma$ consistently
in the bag model by using the MRE method. It was found that $\sigma$ in the pasta phases
slightly decreases with increasing density, and its value lies in the range
of $43-48$~MeV/fm$^{2}$. The dominant contribution to the surface tension
comes from the $s$ quark, which is about one order higher than those from $u$ and $d$ quarks.

We have investigated the hadron-quark mixed phase using the EM method, where the equilibrium
conditions for coexisting phases are derived by minimization of the total energy
including surface and Coulomb contributions. Due to these finite-size effects,
some additional terms appear in the equilibrium conditions for pressures and
chemical potentials, which are different from the Gibbs conditions.
It was found that including the finite-size effects could delay the onset of the
hadron-quark mixed phase and shrink its density range significantly.
On the other hand, the results of pasta phases are very sensitive to the bag
constant $B$ of quark matter. By using a consistent value of $B^{1/4}=210.854$~MeV
in the QMC($L$=40) model and quark matter, a structured mixed phase could be formed in
the density range of $0.70 - 1.47\, \rm{fm}^{-3}$. When $B^{1/4}=180$~MeV is
adopted for quark matter, the density range of the mixed phase is reduced
to $0.43 - 0.80\, \rm{fm}^{-3}$.
It was observed that the results obtained in the EM method lie between those of
the Gibbs and Maxwell constructions.

We examined the influence of nuclear symmetry energy on the hadron-quark phase
transition by using the QMC models with different slope parameter $L$.
It was found that the onset densities of the hadron-quark mixed phase obtained
with QMC($L$=100) are obviously smaller than those with QMC($L$=40),
but there is no visible difference at the end of the mixed phase.

We applied the EOS including the hadron-quark phase transition to study the
properties of neutron stars. In the present calculations, a pure hadronic EOS
of the QMC($L$=40) model predicts a maximum neutron-star mass of $2.08 M_\odot$,
while the resulting radius and tidal deformability of a canonical $1.4 M_\odot$
neutron star are $R_{1.4}= 12.43$ km and $\Lambda_{1.4} = 450$, respectively.
These results are compatible with current constraints from astrophysical observations.

However, the QMC($L$=100) model results in rather large radii and tidal deformabilities
of neutron stars, which are disfavored by the constraints from GW170817.
The recent gravitational-wave event GW190814 has triggered many efforts to explore
the possibility of existing a super-massive neutron star of $\approx 2.6 M_\odot$.
Within the QMC model used in the present work, it is unlikely that such massive
neutron star can be stable. It may be possible to raise the maximum
neutron-star mass by introducing nonlinear self-couplings of the meson fields in
the QMC model, as discussed in the RMF approach~\cite{Fatt20}.
This possibility will be explored in future studies.

When the deconfinement phase transition is included, it is found that using the EM method
with $B^{1/4}=210.854$~MeV and QMC($L$=40), the hadron-quark pasta phases could be formed
in the interior of massive stars with $M > 1.94 M_\odot$, but no pure quark matter exists.
Meanwhile, the maximum neutron-star mass is reduced to $1.96  M_\odot$ in this case,
while the properties of a canonical $1.4 M_\odot$ neutron star remain unchanged.
We emphasize that although the hadron-quark mixed phases do not appreciably change the
neutron-star bulk properties, they could be important for studying cooling observations.
If a small bag constant $B^{1/4}=180$~MeV is adopted for quark matter,
the maximum mass is reduced to $1.49 M_\odot$ in the QMC($L$=40) model,
which is much lower than the constraint of $2 M_\odot$.
This implies that the early onset of the mixed phase with $B^{1/4}=180$~MeV
for quark matter is disfavored due to its large reduction
of the maximum neutron-star mass.
In this work, we used a simple model for quark matter and neglected
possible interactions between quarks, which need to be investigated in future studies.

\section*{Acknowledgment}

This work was supported in part by the National Natural Science Foundation of
China (Grants No. 11675083 and No. 11775119).



\begin{thebibliography}{99}

\bibitem{Baym18} G. Baym, T. Hatsuda, T. Kojo, P. D. Powell, Y. Song, and T. Takatsuka,
Rept. Prog. Phys. \textbf{81}, 056902 (2018).

\bibitem{Glen01} N. K. Glendenning,
Phys. Rep. \textbf{342}, 393 (2001).

\bibitem{Webe05} F. Weber,
Prog. Part. Nucl. Phys. \textbf{54}, 193 (2005).

\bibitem{Demo10} P. B. Demorest, T. Pennucci, S. M. Ranson, M. S. E. Roberts,
and J. W. T. Hessels, Nature (London) \textbf{467}, 1081 (2010).

\bibitem{Fons16} E. Fonseca \textit{et al.}, Astrophys. J. \textbf{832}, 167 (2016).

\bibitem{Arzo18} Z. Arzoumanian \textit{et al.}, Astrophys. J. Suppl. \textbf{235}, 37 (2018).

\bibitem{Anto13} J. Antoniadis \textit{et al.}, Science \textbf{340}, 6131 (2013).

\bibitem{Crom19} H. T. Cromartie \textit{et al.}, Nat. Astron. \textbf{4}, 72 (2020).

\bibitem{Abbo17} B. P. Abbott \textit{et al.}
(LIGO Scientifc Collaboration and Virgo Collaboration),
Phys. Rev. Lett. \textbf{119}, 161101 (2017).

\bibitem{Abbo18} B. P. Abbott \textit{et al.}
(LIGO Scientifc Collaboration and Virgo Collaboration),
Phys. Rev. Lett. \textbf{121}, 161101 (2018).

\bibitem{Abbo19} B. P. Abbott \textit{et al.}
(LIGO Scientifc Collaboration and Virgo Collaboration),
Phys. Rev. X \textbf{9}, 011001 (2019).

\bibitem{Tews18} I. Tews, J. Margueron, and S. Reddy,
Phys. Rev. C \textbf{98}, 045804 (2018).

\bibitem{De18} S. De, D. Finstad, J. M. Lattimer, D. A. Brown, E. Berger, and C. M. Biwer,
Phys. Rev. Lett. \textbf{121}, 091102 (2018).

\bibitem{Fatt18} F. J. Fattoyev, J. Piekarewicz, and C. J. Horowitz,
Phys. Rev. Lett. \textbf{120}, 172702 (2018).

\bibitem{Mali18} T. Malik, N. Alam, M. Fortin, C. Provid\^{e}ncia, B. K. Agrawal, T. K. Jha,
B. Kumar, and S. K. Patra, Phys. Rev. C \textbf{98}, 035804 (2018).

\bibitem{Zhu18} Z. Y. Zhu, E. P. Zhou, and A. Li,
Astrophys. J. \textbf{862}, 98 (2018).

\bibitem{Abbo190425} B. P. Abbott \textit{et al.}
(LIGO Scientifc Collaboration and Virgo Collaboration),
Astrophys. J. Lett. \textbf{892}, L3 (2020).

\bibitem{Abbo190814} B. P. Abbott \textit{et al.}
(LIGO Scientifc Collaboration and Virgo Collaboration),
Astrophys. J. Lett. \textbf{896}, L44 (2020).

\bibitem{Rile19} T. E. Riley \textit{et al.},
Astrophys. J. Lett. \textbf{887}, L21 (2019).

\bibitem{Mill19} M. C. Miller \textit{et al.},
Astrophys. J. Lett. \textbf{887}, L24 (2019).

\bibitem{Glen92} N. K. Glendenning, Phys. Rev. D \textbf{46}, 1274 (1992).

\bibitem{Sche99} K. Schertler, S. Leupold, and J. Schaffner-Bielich,
Phys. Rev. C \textbf{60}, 025801 (1999).

\bibitem{Sche00} K. Schertler, C. Greiner, J. Schaffner-Bielich,
and M. H. Thoma, Nucl. Phys. \textbf{A677}, 463 (2000).

\bibitem{Latt00} A. W. Steiner, M. Prakash, and  J. M. Lattimer,
Phys. Lett. B \textbf{486}, 239 (2000).

\bibitem{Burg02}  G. F. Burgio, M. Baldo, P. K. Sahu, and H.-J. Schulze,
Phys. Rev. C \textbf{66}, 025802 (2002).

\bibitem{Mene03}  D. P. Menezes and C. Provid\^{e}ncia,
Phys. Rev. C \textbf{68}, 035804 (2003).

\bibitem{Yang08}  F. Yang and H. Shen,
Phys. Rev. C \textbf{77}, 025801 (2008).

\bibitem{Xu10} J. Xu, L. W. Chen, C. M. Ko, and B. A. Li,
Phys. Rev. C \textbf{81}, 055803 (2010).

\bibitem{Chen13} H. Chen, G. F. Burgio, H.-J. Schulze, and N. Yasutake,
Astron. Astrophys. \textbf{551}, A13 (2013).

\bibitem{Orsa14} M. Orsaria, H. Rodrigues, F. Weber, and G. A. Contrera,
Phys. Rev. C \textbf{89}, 015806 (2014).

\bibitem{Wu17} X. H. Wu and H. Shen, Phys. Rev. C 96, 025802 (2017).

\bibitem{Bhat10} A. Bhattacharyya, I. N. Mishustin, and W. Greiner,
J. Phys. G \textbf{37}, 025201 (2010).

\bibitem{Wu19} X. H. Wu and H. Shen, Phys. Rev. C 99, 065802 (2019).

\bibitem{Heis93} H. Heiselberg, C. J. Pethick, and E. F. Staubo,
Phys. Rev. Lett. \textbf{70}, 1355 (1993).

\bibitem{Endo06} T. Endo, T. Maruyama, S. Chiba, and T. Tatsumi,
Prog. Theor. Phys. \textbf{115}, 337 (2006).

\bibitem{Maru07} T. Maruyama, S. Chiba, H.-J. Schulze, and T. Tatsumi,
Phys. Rev. D \textbf{76}, 123015 (2007).

\bibitem{Yasu14} N. Yasutake, R. {\L}astowiecki, S. Beni{\'{c}}, D. Blaschke, T. Maruyama, and T. Tatsumi,
Phys. Rev. C \textbf{89}, 065803 (2014).

\bibitem{Spin16} W. M. Spinella, F. Weber, G. A. Contrera, and M. G. Orsaria,
Eur. Phys. J. A \textbf{52}, 61 (2016).

\bibitem{Tats03} T. Tatsumi, M. Yasuhira, and D. Voskresensky,
Nucl. Phys. A \textbf{718}, 359 (2003).

\bibitem{Bao14b} S. S. Bao, J. N. Hu, Z. W. Zhang, and H. Shen,
Phys. Rev. C \textbf{90}, 045802 (2014).

\bibitem{Baym71} G. Baym, H. A. Bethe, and C. J. Pethick,
Nucl. Phys. A \textbf{175}, 225 (1971).

\bibitem{Latt85} J. M. Lattimer, C. J. Pethick, D. G. Ravenhall, and D. Q. Lamb,
Nucl. Phys. A \textbf{432}, 646 (1985).

\bibitem{Latt91} J. M. Lattimer and F. D. Swesty,
Nucl. Phys. A \textbf{535}, 331 (1991).

\bibitem{Guic88} P. A. M. Guichon,
Phys. Lett. B {\bf 200}, 235 (1988).

\bibitem{Sait96} K. Saito, K. Tsushima, and A. W. Thomas,
Nucl. Phys. A {\bf 609}, 339 (1996).

\bibitem{Mull98} H. Muller and B. K. Jennings,
Nucl. Phys. A {\bf 640}, 55 (1998).

\bibitem{Yue06} P. Yue and H. Shen,
Phys. Rev. C \textbf{74}, 045807 (2006).

\bibitem{Sait07} K. Saito, K. Tsushima, and A. W. Thomas,
Prog. Part. Nucl. Phys. {\bf 58}, 1 (2007).

\bibitem{Guic18} P. A. M. Guichon, J. R. Stone, A. W. Thomas,
Prog. Part. Nucl. Phys. {\bf 100}, 262 (2018).

\bibitem{Gram17} G. Grams, A. M. Santos, P. K. Panda, C. C. Provid\^{e}ncia, D. P. Menezes,
Phys. Rev. C \textbf{95}, 055807 (2017).

\bibitem{Mott19} T. F. Motta, A. M. Kalaitzis, S. Antic,
P. A. M. Guichon, J. R. Stone, A. W. Thomas,
Astrophys. J. \textbf{878}, 159 (2019).

\bibitem{Panda04a} P. K. Panda, D. P. Menezes, and C. Provid\^{e}ncia,
Phys. Rev. C \textbf{69}, 025207 (2004).

\bibitem{Panda04b} P. K. Panda, D. P. Menezes, and C. Provid\^{e}ncia,
Phys. Rev. C \textbf{69}, 058801 (2004).

\bibitem{Panda10} P. K. Panda, C. Provid\^{e}ncia, and D. P. Menezes,
Phys. Rev. C \textbf{82}, 045801 (2010).

\bibitem{Panda12} P. K. Panda, A. M. S. Santos, D. P. Menezes, and C. Provid\^{e}ncia,
Phys. Rev. C \textbf{85}, 055802 (2012).

\bibitem{Berg87} M. S. Berger and R. L. Jaffe,
Phys. Rev. C \textbf{35}, 213 (1987); \textbf{44} 566(E) (1991).

\bibitem{Palh10} L. F. Palhares and E. S. Fraga,
Phys. Rev. D \textbf{82}, 125018 (2010).

\bibitem{Pint12} M. B. Pinto, V. Koch, and J. Randrup,
Phys. Rev. C \textbf{86}, 025203 (2012).

\bibitem{Lugo17} G. Lugones and A. G. Grunfeld,
Phys. Rev. C \textbf{95}, 015804 (2017).

\bibitem{Lugo19} G. Lugones and A. G. Grunfeld,
Phys. Rev. C \textbf{99}, 035804 (2019).

\bibitem{Alfo01} M. G. Alford, K. Rajagopal, S. Reddy, and F. Wilczek,
Phys. Rev. D \textbf{64}, 074017 (2001).

\bibitem{Lugo13} G. Lugones, A. G. Grunfeld, and M. A. Ajmi,
Phys. Rev. C \textbf{88}, 045803 (2013).

\bibitem{Bao15} S. S. Bao and H. Shen,
Phys. Rev. C \textbf{91}, 015807 (2015).

\bibitem{Oert17} M. Oertel, M. Hempel, T. Kl\"{a}hn, and S. Typel,
Rev. Mod. Phys. \textbf{89}, 015007 (2017).

\bibitem{LiBA08} B. A. Li, L. W. Chen, and C. M. Ko,
Phys. Rep. \textbf{464}, 113 (2008).

\bibitem{Tews17}  I. Tews, J. M. Lattimer, A. Ohnishi, and E. E. Kolomeitsev,
Astrophys. J. \textbf{848}, 105 (2017).

\bibitem{Zhan13} Z. Zhang and L. W. Chen, Phys.
Lett. \textbf{B726}, 234 (2013).

\bibitem{BPS71} G. Baym, C. Pethick, and P. Sutherland,
Astrophys. J. \textbf{170}, 299 (1971).

\bibitem{Ji19} F. Ji, J. N. Hu, S. S. Bao, and H. Shen,
Phys. Rev. C \textbf{100}, 045801 (2019).

\bibitem{Hind08} T. Hinderer, Astrophys. J. \textbf{677}, 1216 (2008);
\textbf{697}, 964(E) (2009).

\bibitem{Post10} S. Postnikov, M. Prakash, and J. M. Lattimer,
Phys. Rev. D \textbf{82}, 024016 (2010).

\bibitem{Fatt20} F. J. Fattoyev, C. J. Horowitz, J. Piekarewicz, and B. Reed,
Phys. Rev. C \textbf{102} 065805 (2020).

\end{thebibliography}
\end{document}